\documentclass[fleqn,usenatbib]{mnras}

\usepackage{newtxtext,newtxmath}

\usepackage[T1]{fontenc}

\DeclareRobustCommand{\VAN}[3]{#2}
\let\VANthebibliography\thebibliography
\def\thebibliography{\DeclareRobustCommand{\VAN}[3]{##3}\VANthebibliography}

\usepackage{graphicx}	
\usepackage{amsmath}	
\usepackage{bm}

\def\ga{\Gamma}
\def\mdd{\dot{M}}

\def\ie{{\em i.e., }}
\def\vp{v_{\rm po}}

\def\vfi{v_{\phi}}
\def\bp{B_{\rm po}}
\def\bfi{B_{\phi}}
\def\ap{A_{\rm po}}

\def\rg{r_{\rm g}}
\def\rd{r_{\rm d}}
\def\rs{r_{\rm s}}

\def\rc{r_{\rm c}}

\def\te{T_{\rm e}}
\def\tp{T_{\rm p}}
\def\tpm{T_{\rm pin|max}}
\def\ne{n_{\rm e}}
\def\np{n_{\rm p}}

\def\thetae{\Theta_{\rm e}}
\def\thetap{\Theta_{\rm p}}

\def\shokst{{\cal S}}
\def\ebb{\mathbb{E}}
\def\pbb{\mathbb{P}}
\def\lsim{\lower.5ex\hbox{$\; \buildrel < \over \sim \;$}}
\def\gsim{\lower.5ex\hbox{$\; \buildrel > \over \sim \;$}}

\def\msol{M_{\odot}}
\def\mdot{\dot{M}}
\def\rg{r_{\rm g}}
\def\gamp{\Gamma_{\rm p}}
\def\game{\Gamma_{\rm e}}
\def\qem{Q_{\rm e}^-}
\def\qep{Q_{\rm e}^+}
\def\vvec{\bm{v}}
\def\vpvec{\bm{v}_{\rm po}}
\def\bvec{\bm{B}}
\def\bpvec{\bm{B}_{\rm po}}
\def\ephivec{\bm{\hat e}_\phi}
\def\rtil{\tilde{r}}
\def\qplus{Q^{+}}
\def\qminus{Q^{-}}

\def\rco{{r}_{\rm co}}
\def\rhoe{\rho_{\rm e}}
\def\qcc{Q_{\rm cc}}
\def\qbr{Q_{\rm br}}
\def\qsyn{Q_{\rm syn}}
\def\qcsy{Q_{\rm csy}}
\def\qcbr{Q_{\rm cbr}}
\def\qbb{Q_{\rm bb}}
\def\qcbb{Q_{\rm cbb}}

\newcommand{\nel}{n_{\rm e}}

\newcommand{\me}{m_{\rm e}}
\newcommand{\mpr}{m_{\rm p}}
\newcommand{\nh}{n_{\rm in}}
\newcommand{\neh}{n_{\rm e in}}
\newcommand{\nph}{n_{\rm p in}}
\newcommand{\feh}{f_{\rm e in}}
\newcommand{\fph}{f_{\rm p in}}
\newcommand{\theh}{\Theta_{\rm ein}}
\newcommand{\thph}{\Theta_{\rm pin}}
\newcommand{\tph}{T_{\rm p in }}

\newcommand{\polyp}{N_{\mbox{{\scriptsize p}}}}
\newcommand{\polye}{N_{\mbox{{\scriptsize e}}}}

\newcommand{\mdotin}{{\dot{\mathcal{M}}}_{\rm{in}}}
\newcommand{\rin}{r_{\rm in}}
\newcommand{\vin}{v_{\rm in}}
\newcommand{\thetain}{\Theta_{\rm in}}

\newcommand{\rps}{r_{\rm ps}}
\newcommand{\rss}{r_{\rm ss}}

\newcommand{\rci}{r_{\rm c}^{\rm in}}
\newcommand{\rcm}{r_{\rm c}^{\rm mid}}
\newcommand{\rcou}{r_{\rm c}^{\rm out}}

\newcommand{\as}{a_{\rm s}}
\newcommand{\rsp}{r_{\rm ps}}

\usepackage{tabularx}

\defcitealias{sc18a}{KSC18a}
\defcitealias{sc18b}{KSC18b}
\defcitealias{sc19a}{SC19a}
\defcitealias{sc19b}{SC19b}
\defcitealias{scp20}{SCL20}
\defcitealias{sc22}{SC22}

\title[Funnel flows]{Two-temperature accretion flows around strongly magnetized stars and their spectral analysis}

\author[Sarkar et al.]{
Shilpa Sarkar,$^{1,2,5}$\thanks{E-mail: shilpa.sarkar30@gmail.com}
Kuldeep Singh,$^{1}$
Indranil Chattopadhyay$^{1}$\thanks{E-mail: indra@aries.res.in}
and Philippe Laurent$^{3,4}$
\\
$^{1}$Aryabhatta Research Institute of Observational Sciences 
(ARIES), Manora Peak, Nainital, Uttarakhand 263002, India\\
$^{2}$Pt. Ravishankar Shukla University, Great Eastern Rd, Amanaka, Raipur, Chhattisgarh 492010, India\\
$^{3}$IRFU / Service d'Astrophysique, Bat. 709 Orme des Merisiers, CEA Saclay, 91191 Gif-sur-Yvette, Cedex France\\
$^{4}$Laboratoire Astroparticule et Cosmologie, B\^atiment Condorcet, 10, rue Alice Domont et L\'eonie Duquet, 75205 Paris, Cedex France \\
$^{5}$IUCAA, Ganeshkhind, Pune 411007, India
}

\date{Accepted XXX. Received YYY; in original form ZZZ}

\pubyear{2015}

\begin{document}
\label{firstpage}
\pagerange{\pageref{firstpage}--\pageref{lastpage}}
\maketitle

\begin{abstract}
We investigate two-temperature accretion flows onto strongly magnetized compact stars. 
Matter is accreted in the form of an accretion disc upto the disc radius ($\rd$), 
where, the magnetic pressure exceeds both the gas and ram pressure and thereafter the matter is channelled along the field lines onto the poles.
We solve the equations of motion self-consistently along the field lines, incorporating radiative processes like bremsstrahlung, synchrotron and inverse-Comptonization. 
For a given set of constants of motion, the equations of motion do not produce unique transonic solution.  
Following the second law of thermodynamics the solution with the highest entropy is selected and thereby eliminating the degeneracy in solution. 
We study the properties of these solutions and obtain corresponding spectra as a function of the magnetic field ($B_*$), spin period ($P$) and accretion rate of the star ($\mdot$).
A primary shock is always formed just near the surface. The enhanced radiative processes in this post-shock region slows down the matter and it finally settles on the surface of the star. 
This post-shock region contributes to $\gtrsim 99.99\%$ of the total luminosity obtained from the accretion flow. It is still important to study the full accretion flow
because secondary shocks may be present for some combination of $B_*$, $P$ and $\mdot$ in addition to primary shocks. We find that secondary shocks, if present, produce an extended emission at higher energies in the spectra. 
\end{abstract}

\begin{keywords}
stars -- accretion flows -- shocks -- magnetic fields -- spectra
\end{keywords}



\section{Introduction}
Accretion onto magnetized stars remains one of the hot topics of research interest since the discovery of X-ray pulsars \citep{gia71}. 
%
These pulsars were later found to be neutron stars (NSs) accreting matter from their binary counterpart, either via stellar wind or Roche lobe overflow \citep{pr72,do73, lam73}. {The matter accreted generally possess some angular momentum because of which it first forms an accretion disc similar to a black hole (BH) accretion disc.} 
{
{But in the presence of strong magnetic field, accretion disc terminates at some radius called the magnetospheric radius.  Thereafter, accretion  proceeds along the curved magnetic field lines. \cite{pr72} suggested that emission would mostly come from a region close to the poles of the star. }

{\cite{kol02} studied the behaviour of accretion flow along an aligned dipole magnetic field  around rotating magnetized stars (NSs and
young stellar objects), \ie along a curved flow geometry. They  utilised the magneto-hydrodynamic (MHD) integrals \citep{wd67, m68, lo86} as well as Bernoulli parameter that are conserved along the field lines, to obtain {global} transonic accretion solution, { connecting the accretion disc to the poles of the star. }
 It is important to note that since stars possess a hard surface, matter accreted should settle down onto the surface of the star  \citep{fukue87}. This demands the formation of a shock, whereby the kinetic energy of the matter could be radiated away \citep{li96}. 
 {\cite{kol02} did not address this issue and considered only adiabatic flows which possessed supersonic velocity near the star's surface}. {\cite{ka08} followed \citeauthor{kol02} and obtained shocked solutions, however the shock obtained was  located farther from the star's surface, while it is more likely that the terminating shock for the accretion column should be nearer to the star's surface. Moreover, in their work, even  the post-shock flows have relatively higher speeds close to the star's surface, \ie did not satisfy the star's surface boundary conditions. }
Since the accretion column is expected to be terminated at a shock close to the surface, thus a large number of works focused just on the region close to the post-shock accretion column, which enabled them to study in detail the emission processes {responsible for} the observable spectrum \citep{d73,arons78,beck98,bw05b,bw05a,bw07,beck12,bw19,bw20}.}

{\cite{sc18a} followed the methods of \cite{kol02,ka08} but included  cyclotron and bremsstrahlung cooling. They obtained self-consistent magnetized accretion solutions, which connected the flow from the inner region of the
accretion disc ($\rd$) to the star's pole via a surface shock (also known as primary shock, $\rps$). 
They discussed the importance of radiative cooling in the post-shock flow for the matter to  slow down with asymptotically zero speed near the star's surface. } 
{However, the temperatures of an accretion flow starting from the accretion disc edge to the poles of the star varies by more than 3-4 orders of magnitude. For such a wide variation in temperature, a fixed $\Gamma$ (i. e., adiabatic index) equation of state (EoS) for the gas is untenable \citep{t48,c38,rcc06}. \cite{sc18b} extended their previous work  and instead of using a fixed $\Gamma$ EoS, they now used a variable $\Gamma$ EoS proposed by \cite{cr09}, also known as CR EoS, which is dependent on the temperature as well as the composition of the flow. \cite{sc18b} showed that there are  multiple sonic point regions in magnetized accretion flows. Therefore, many modes of accretion through the bipolar magnetic field funnels are possible, and not just a column or a conical flow onto an optically thick post-shock region. }

\subsection*{{Two-temperature accretion flows}}

{Ionized astrophysical plasma is composed of different particles (electrons and protons). If these particles are not given} sufficient time to interact within themselves, or technically, if
the Coulomb coupling between the species is weak, {then} this would lead to a two-temperature flow, {where protons and electrons would be defined by two different temperature distributions}. In most of the astrophysical systems, this condition is {found to be} valid. The infall timescales are generally very much shorter than the Coulomb coupling timescales \citep{sle76,s83,cmt84,p90,yn14}.
{In addition, 
radiation mechanisms acting on electrons and {protons are different because of their different masses} and scattering cross-sections. 
Thus, electrons and protons are likely to settle down into two different temperature distributions.}  
{A lot of work has been done in two-temperature accretion flows around BHs \citep{naka96,man97,fn03,samir05,sado16,sadoetal17,indu18,indu20,c18}, as well as for flows around magnetized stars \citep{ss75,lr82,ny95,sax05,beck17a,beck17b,bu20,bw22}.
}

\cite{lt80} (hereafter, LT80) identified a problem while solving two-temperature solutions around compact objects. With respect to one-temperature {flows, we have} an additional variable in two-temperature system which is the extra temperature. To obtain a solution the above {authors made} an arbitrary assumption. To quote LT80
verbatim ``..... because of the uncertainty in the mechanism coupling electrons and ions, we simply
parameterize $\tp/\te$ as a constant". In other words, it suggests that if the ratio between the temperatures is changed to some other constant value, we would obtain a completely different solution. This indicated that there is a degeneracy present in the two-temperature system unlike in case of one-temperature flows where for a given set of CoM we get a unique transonic solution.  This degeneracy is irrespective of the type of central object and is generic to two-temperature flows. {Apart from LT80, there are a number of papers indicating the same degeneracy issue. Similar to LT80 where $\tp/\te$ is parameterized, other works followed some other methodology to constrain the degeneracy. We have grouped them together and have discussed them below. We note that the works discussed below are related to magnetized stars only.}

\noindent {\textit{Parameterising of shock jump values:} In {1975, \citeauthor{ss75}} 
studied funnel flows in two temperature regime, but considered only the 
post-shock region to obtain solutions and {{compute}} the spectrum. They considered the ratio between $\tp$ and $\te$  just after the shock as a parameter (marked as $\beta_{\rm s}$ in their paper). Similar approach was adopted by \cite{sax05} for obtaining two-temperature accretion solutions around white dwarfs and \cite{beck17a,beck17b} used it for NSs.}

\noindent {\textit{Assumption of additional relation to determine $\tp$ and $\te$:}
\cite{lr82} also considered the post shock accretion column while computing the spectrum. However, they utilised an arbitrary assumption to obtain the value of $\te$. 
They assumed that the heating of electrons by ions ($\qep$) equals the radiative cooling ($\qem$). In other words, $\qep=\qem$. This assumption is arbitrary and need not be true, since some amount of electron heat could be advected inwards with the flow towards the central object \citep[in case of BHs]{man97}. A similar approach was used by \cite{mn01} where they obtained full global but self-similar solutions around an NS utilising the above assumption with weak magnetic field. }

In a series of 
papers by \cite{sc19a,sc19b} \citepalias[hereafter,][]{sc19a,sc19b}, \cite*{scp20}
\citepalias[hereafter,][]{scp20} and \cite{sc22} \citepalias[hereafter,][]{sc22} which are based on two-temperature accretion flows around BHs, it has been discussed that these 
flows are degenerate in nature. The reason for this degeneracy is the increase in the number of {flow} variables (an extra temperature) without any increase in the number of requisite equations. 
{In one-temperature regime, however, this was not the case and for a given set of CoM, a unique transonic solution existed (also, see \citetalias[][]{sc18a,sc18b}). But in the two-temperature regime, 
infinite number of transonic solutions were \{obtained}. 
In the papers \citetalias{sc19a,sc19b,scp20,sc22}, this problem of degeneracy was attended and a novel methodology to 
constrain it was proposed. Apart from energy and other CoM, fluids are also characterized by  entropy. 
\citeauthor{b52} in 1952 concluded that a transonic solution is the one {with maximum} entropy and would be  preferred by nature. Also, \cite{bl03,bsl08} used the concept of entropy close to the BH horizon, in addition to other integrals of motion,  to obtain a transonic BH accretion solution. Although these works were done in the one-temperature regime, but the concept provided by these authors, that entropy {can be used as a tool to chose the correct solution, also} served as the basis to remove  degeneracy in two-temperature solutions. 
The integration of the first law of thermodynamics gives the measure of entropy. In two-temperature regime, we have two differential equations for temperature ($d\tp/dr$ and $d\te/dr$) coupled by the Coulomb coupling term which inhibits the integration of these equations to obtain an analytical expression of entropy. Fortunately the presence of event horizon in case of BH solves the problem. Utilising the fact that close to the event horizon, gravity overpowers any other interaction or processes and matter velocities approach free-fall velocities, the first law of thermodynamics can be integrated and
an expression for entropy   is admissible, which is strictly valid near the horizon. Using this entropy measure, \citetalias{sc19a,sc19b,scp20,sc22} obtained unique
two-temperature accretion solutions around BHs.

The situation gets complicated for the case of magnetized stars which possess a hard surface unlike the BH event horizon. 
Thus the form of entropy measure proposed above  cannot be applied {in a similar fashion for}  flows around magnetized stars. 
Although the gravity of magnetized stars  makes the accreting matter supersonic, but the presence of hard surface drives a
shock in the accretion column \citep[see also][]{li96}, after which the velocity of the matter reduces to negligible values.
Even if it is assumed that the supersonic matter directly hits the star's surface 
\citep{kol02} without forming a shock, the matter would still not achieve velocities high enough ($v\nsim c$), to use the entropy expression and obtain a
measure of entropy close to the surface. 
This leads to a serious problem of constraining the degeneracy in two-temperature accretion flows around magnetized stars. Therefore, in this paper, 
we propose a novel methodology to remove degeneracy and obtain unique transonic two-temperature accretion solutions
around magnetized stars for a given set of CoM. We elaborately discuss it in the methodology section. 
Thereafter, we investigate accretion solutions for a large set of parameter space, to get a 
global picture of these accreting systems. In addition, we also perform the spectral analysis. 

This paper is arranged according to the following sections. In section \ref{sec:basic_equations}, we introduce the basic equations
and assumptions used and in section \ref{sec:meth}, we discuss the methodology to obtain a unique transonic two-temperature solution. 
In section \ref{sec:result}, we present and discuss the results obtained for a large set of parameter space. We also present spectral 
analysis in this section and then conclude in section \ref{sec:con}.


$  $\section{Basic Equations of Motion and Assumptions}
\label{sec:basic_equations}
\subsection{MHD equations}
\label{sec:mhdeq.eq}
The work is done in the ideal MHD regime assuming steady, axisymmetric and inviscid flow \citep{c56,h78,lo86,u99} in spherical
coordinate system ($r,~\theta,~\phi$). The velocity and magnetic field are given by, $\vvec=\vpvec+ v_\phi \ephivec$ and
$\bvec=\bpvec+ B_\phi \ephivec$, respectively. The subscripts {`po'}  and $\phi$ represents the poloidal and toroidal component
respectively and $\bm{\hat e}$ is the unit vector. In steady state and under axisymmetry assumption, the basic MHD equations are as follows :
\begin{align}
&{\rm Mass~conservation~equation~:~}\nabla\ldotp(\rho \vvec) = 0, \label{eq:cont} \\
&{\rm Momentum~conservation~equation~:~}\nonumber\\
&~~~~~~~~~~~~~~~~~~~~~~~~~~~~~~~~~~~~(\rho\vvec\ldotp\nabla)\vvec = - \nabla p + \frac{1}{c}(\bm{J}\times\bm{B})- \rho\nabla\Phi_{\rm g}, \label{eq:euler} \\
&{\rm Faraday's~law~using~ideal~Ohm's~law:~}\nabla \times (\vvec \times \bm{B}) =0,\label{eq:faraday}\\
&{\rm Divergence~constraint~:~}\nabla\cdot\bm{B} = 0, \label{eq:divb} 
\end{align}
where, $\rho$ is the mass density, $p$ is the isotropic plasma pressure, $c$ is the speed of light, $\Phi_{\rm g}$ is the
gravitational potential of the star and $\bm{J}$ is the current density which is $=(\nabla \times {\bm B})c/4\pi$  from Ampere's law.
In order to mimic the  effects of strong gravity we use the \citeauthor{pw80} potential \citep{pw80} throughout this work, which is 
given by, $\Phi_{\rm g}=-GM_*/(r-\rg)$ where $\rg=2GM_*/c^2$ is the Schwarzschild radius, $G$ is the gravitational constant and 
$M_*$ is mass of the star. 

Apart from the above equations, we need the first law of thermodynamics to study the temperature variation inside the system in 
the presence of advection and dissipation, which is given by :
\begin{equation}
\frac{p}{\rho^2}\frac{d\rho}{ds}-\frac{d({e}/\rho)}{ds}= \frac{\qplus - \qminus}{\rho \vp} = \frac{\Delta Q}{\rho \vp},
\label{eq:flt}
\end{equation}
where, ${e}$ is the internal energy density, $\qplus$ and $\qminus$ are the heating and cooling rates, respectively and 
$\Delta Q=\qplus-\qminus$. The first law of thermodynamics is written separately for the two different species (protons and electrons), 
but are coupled by the Coulomb coupling term, which is responsible for energy exchange  between the protons and electrons. Detailed 
discussion regarding the dissipative processes is given in section \ref{sec:rad}.

\begin{table} 
\caption{{The variables and some abbreviations used in the paper and their description}}
\label{table:2}
\centering
\begin{tabular}{p{6.em}| p{22em}}
\hline\hline
{Variable}&{Description} \\
\hline
{$v$}& {Flow velocity (subscript $\phi$ $\rightarrow$ toroidal component, po $\rightarrow$ poloidal component)}\\
{$B$}& {Magnetic field (subscript definition same as $v$)}\\
$\rho$ & {Mass density} \\
$\Phi$& {Potentials (subscript g $\rightarrow$ Gravitational potential, centri $\rightarrow$ potential due to centrifugal forces)}\\
$p$ & {Pressure} \\
$e$& {internal energy density} \\
$Q^+$, $Q^-$ & {Heating and cooling rates} \\
$ds$ & {Differential line element along the field line}\\
$\dot{M}$, $\dot{M}_{\rm Edd}$ & {Accretion rate, Eddington rate} \\
$\ap$& {Cross-sectional area perpendicular to $\bp$} \\
$\kappa(\Psi)$& {Mass flux to magnetic flux ratio} \\
$\Omega(\Psi)$& {Angular velocity of the field lines} \\
$\omega$& {Angular velocity of the matter} \\
$L(\Psi)$& {Total angular momentum} \\
$E(\Psi)$& {Generalised Bernoulli parameter} \\
${\cal E}$& {Canonical form of Bernoulli constant} \\
$h$& {Specific enthalpy} \\
$r$& {Radius in spherical coordinates} \\
$\theta$& {Co-latitude of $r$} \\
$\Psi$& {Stream function or magnetic flux function} \\
$\mu$& {Magnetic moment} \\
$\rd=\mu/\Psi$ & {Disc radius defined as the radius from the center of the star to the point where the field line crosses the equatorial plane of the disc.} \\
$\rco$& {Co-rotation radius} \\
$\alpha$& {Ratio between the co-rotation radius and disc radius($=\rco/\rd$)}\\
$n_i$ & {Number density ($i=e,~p$ for electrons and protons respectively)}. \\
$m_i$& {Mass of $i^{\rm th}$ species} \\
$T_i$& {Temperature of $i^{\rm th}$ species} \\
$\Theta_i$& {Dimensionless temperature defined w.r.t the rest mass of the species [$=kT_i/(m_ic^2$)]} \\
$\Gamma_i$& {Adiabatic index for i$^{th}$ species} \\
$N_i$& {Polytropic index for i$^{th}$ species} \\
$\Delta Q_i$& {Difference in the heating and cooling rates (=$Q^+_i-Q^-_i$)} \\
$\as$& {Sound speed} \\
$M$& {Mach number ($v/\as$)} \\
$M_*$& {Mass of the NS} \\
$r_*$& {Radius of the NS} \\
$B_*$ & {Surface magnetic field of the NS} \\
$B_{10}$& {$B_*$ in terms of $10^{10}$G} \\
$P$& {Period of the NS} \\
$\rg$& {Schwarzschild radius (=$2GM_*/c^2$)} \\
${\cal \mdot}$& {Entropy accretion rate} \\
$\rin$ & {A point where the fluid velocity reaches free-fall velocity} \\
{Subscript `in'}& {Represents the value of variables ($r, v, n_i,T_i,\Theta_i,f_i, A, \mdot, {\cal M}$) at $r\rightarrow \rin$}\\
$\tpm$& {Proton temperature for maximum entropy solution} \\
{CR}& {Compression ratio at the shock location} \\
{Subscript `c'}& {Represents the value of variables ($r,v,M,T_i,\Theta_i$) at the sonic point} \\
$\rci,\rcm,\rcou$& {Inner, middle and outer sonic points respectively} \\
{Subscript `ps'}& {Represents the value of variables ($r,CR$) at the primary shock location} \\
{Subscript `ss'}& {Represents the value of variables ($r,CR$) at the secondary shock location}\\
{EoS}&  {Equation of state} \\
{EoM}& {Equation of motion} \\
{CoM}& {Constant of motion} \\
{TS} & {Transonic Solution} \\
\hline
\end{tabular}
\end{table}

\subsection{MHD integrals}

We introduce a flux function $\Psi (r,\theta)$ (\citealp{lo86,u99,kol02}; \citetalias{sc18a,sc18b}), which represents a specific magnetic field line. 
On integrating the equations numbered  (\ref{eq:cont}) to (\ref{eq:divb}), we obtain quantities that remain conserved along these field lines. They are :  $\kappa(\Psi)$, $\Omega(\Psi)$, $L(\Psi)$ and $E(\Psi)$ arising due to the conservation of mass, angular velocity of field lines, total angular momentum and energy, respectively. The derivation of these quantities are discussed below. \\
Integrating continuity equation (\ref{eq:cont}), gives the equation for conservation of mass flux, which is given by,
\begin{equation}
 \mdd = \rho \vp\ap = \mbox{constant},
 \label{eq:accrate}
\end{equation}
where, $\mdd$ is known as the accretion rate of the system.
From the Eq.~(\ref{eq:divb}), we obtain the magnetic flux conservation,
\begin{equation}
 \bp\ap = \mbox{constant},
 \label{eq:magflux}
\end{equation}
where, $\ap$ is the cross-sectional area perpendicular to the magnetic field $\bp$. From Eqs.~ 
(\ref{eq:accrate}) and (\ref{eq:magflux}), we obtain a relation between $\vp$, $\rho$ and
$\bp$ which is given by,
\begin{equation}
 \vp = \frac{\kappa(\Psi)}{4\pi\rho}\bp,
 \label{eq:constk}
\end{equation}
where, $\kappa(\Psi)=$ constant, is the mass flux to magnetic flux ratio.\\
{The poloidal field lines are the streamlines, so
a differential element on the field line is}
\begin{equation}
{{ds}^2=dr^2+r^2d\theta^2=dr^2\left[1+r^2\left(\frac{d\theta}{dr}\right)^2\right].}
\label{eq:ds}
\end{equation}
{So, $\bp$ and $\vp$ are locally tangential to ${ds}$.}

The Faraday equation (\ref{eq:faraday}) gives  conservation of the angular velocity $\Omega(\Psi)$ of  field lines,
\begin{equation}
 \Omega\left(\Psi\right) = \omega - \frac{\kappa(\Psi) \bfi}{4\pi\rho \rtil} = \mbox{constant},
 \label{eq:constomega}
\end{equation}
where, $\rtil=r {\rm sin}\theta$ and $\omega = \vfi/\rtil$ is the angular velocity of the matter.\\ 
From the azimuthal component of Euler equation (\ref{eq:euler}), we get the conservation of total angular momentum $L(\Psi)$,
\begin{equation}
L(\Psi) = \omega \rtil^{2} - \frac{\bfi\rtil}{\kappa(\Psi)} = \mbox{constant}.
\label{eq:constang}
\end{equation}  
On integrating the poloidal component of the Euler equation (\ref{eq:euler}) with the help of Eqs.~
(\ref{eq:constomega} \& \ref{eq:flt}), we get the generalized Bernoulli constant or in other words, the total energy $E$ of the flow, which is given by,
\begin{equation}
E(\Psi) = \frac{1}{2}{\vp^{2}} + \frac{1}{2}(\omega-\Omega)^{2}\rtil^{2} + h + \Phi_{\rm g}- \frac{\Omega^{2}\rtil^{2}}{2}+\int\frac{\Delta Q }{\rho \vp}{ds} = \mbox{constant},
\label{eq:bernoulli}
\end{equation}
where, $h$ denotes the enthalpy. The last term arises due to the presence of dissipative processes in the system. {
\subsection{Stream function and the strong magnetic field assumption}
\label{sec:sflassump}
{We assume the magnetized star to have dipole-like magnetic field.  
The magnetic flux function for this field in spherical coordinates is,
\begin{equation}
 {\Psi=\frac{\mu}{r}{\rm sin}^{2}\theta~,}
\label{eq:psi}
\end{equation}
and the geometry of the dipole field lines is given by,
\begin{equation}
 {r=\rd(\Psi){\rm sin}^{2}\theta,}
\label{eq:rd}
\end{equation}
where, { $r$ is the spherical radial coordinate, $\theta$ is the co-latitude}, $\mu$ is the magnetic moment and $\rd=\mu/\Psi$ is the radius from the center of the star to the point where the field line ($\Psi)$ crosses the equatorial plane of
the disc. 
In our work, $\rd$ is also the radius from where the matter starts channelling through the magnetic field lines from the
accretion disc as shown in Fig.~(\ref{fig:1}). We see from this equation that $r$ and $\theta$ are not independent and are constrained through this relation. {The poloidal magnetic field considered is dipolar \ie ${\mathbf{\bp}=3\mathbf{r}(\bm{\mu}\cdot \mathbf{r})/r^5-\bm{\mu}/r^3}$ and can be simplified to get} {\citep{kol02,ka08}}, 
\begin{equation}
{\bp(r)=\frac{\mu}{r^3}\sqrt{4-\frac{3r}{\rd}}.}
\label{eq:bp}
\end{equation} }
{
We assume  the star's rotation axis to be co-aligned with the magnetic 
moment ($\mu$) $\Rightarrow \Omega \parallel \mu$ (see, Fig.~\ref{fig:1}). {Since our main focus in this work is to find consistent two temperature accretion solution onto magnetised stars (not particularly pulsars), we chose the cases where the magnetic field axis and rotation axis are aligned. Simplified as the case may be, but our analysis captures the qualitative features of these types of flows. 
\cite{lm88} showed that old pulsars have aligned axes and this alignment is achieved in times scales of the order of  $10^7$ years \citep{yl23, ba21}. Even some young pulsars with similar alignment are also reported in literature. 
The current  paper directly applies to these types of system,  
and we aim to attend the issue of misalignment in some subsequent work.} In addition, {we assume the star's magnetic field to be strong enough such that the matter flow does not affect the magnetic field geometry}. This is valid when the magnetic energy density is much larger than the gas or ram pressures \citep{kol02} and can be expressed as, 
\begin{equation}
{{\bp^2}/{8\pi}\gg p,~(\rho \vp^{2})}
\label{eq:sbassump}
\end{equation}

In the strong magnetic field limit and after some simple calculations  we arrive at two main conclusions (for the derivation part see, \cite{kol02}, \citetalias{sc18a,sc18b}), which can be represented as:
\begin{equation}
{|{\omega-\Omega}|\ll {\Omega} ~~~~{\rm and}~~~~\frac{\bfi}{\bp}\ll 1.}
\label{eq:ombfi3}
\end{equation}

The first relation implies that matter moves with the same angular velocity as the  field lines. 
Additionally since these field lines are strongly anchored to the surface of the star, they rotate with the same angular velocity as that of the star,  or $\Omega_{\rm matter}=\Omega(\Psi)=\Omega_{\rm star}=\Omega$.
This also implies that $\rd$ is very close to the co-rotation radius ($\rco$) or $\rd \simeq \rco$. Second relation dictates that it is safe to ignore the toroidal component of the magnetic field which is negligibly small compared to its poloidal component.
}

\begin{figure}
\centering
\hspace{0.0cm}
\includegraphics[width=8.8cm,trim={0.cm 0.cm 0.cm 0.cm},clip]{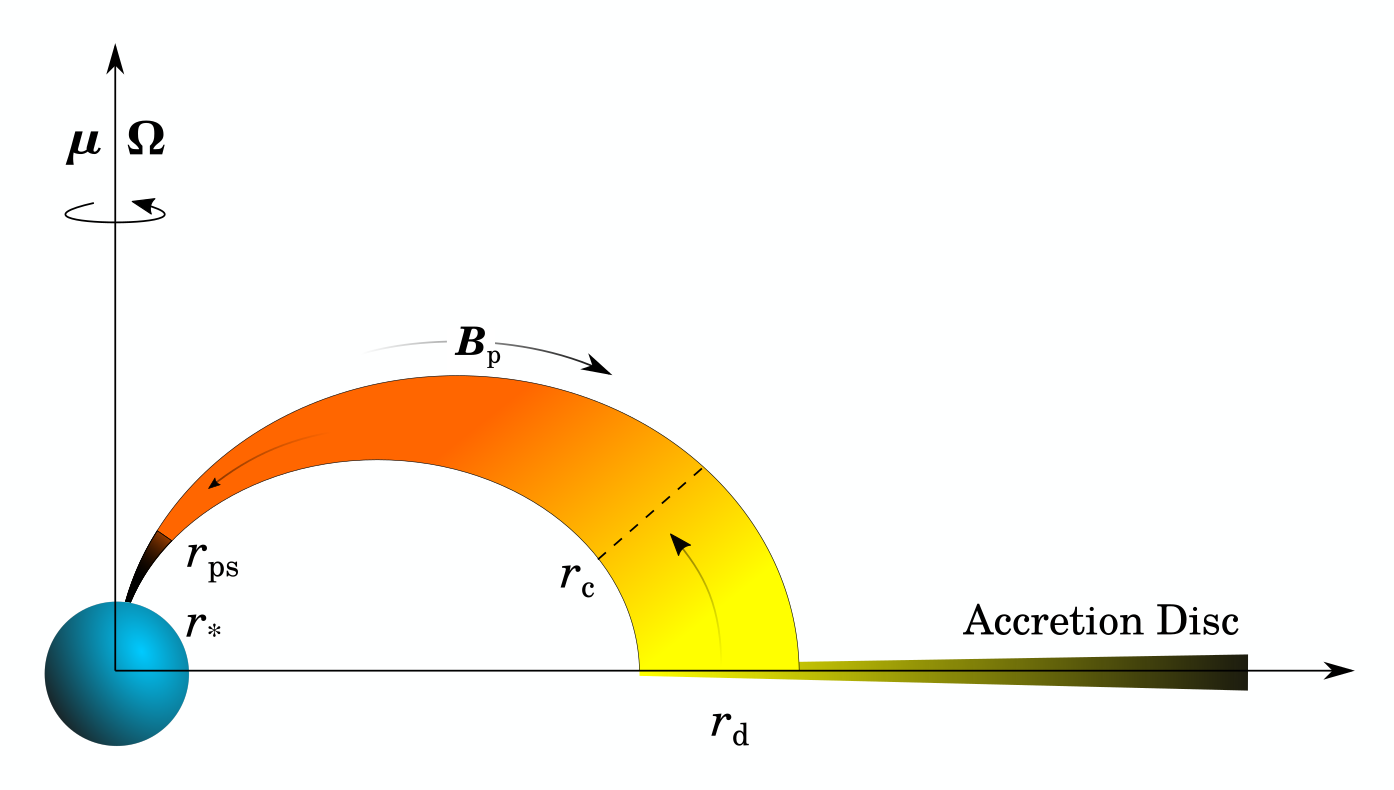}
\vspace{0.0cm}
\caption[] {\small Representation of a magnetized accretion flow. 
The disc radius is marked as $\rd$, sonic point as
$\rc$, 
primary shock as $\rps$ and $r_*$ is the radius of the star. 
}
\label{fig:1}
\end{figure}
{The effective potential of an accretion flow along a given field line $\Psi(r,\theta)=$ constant, in a reference frame co-rotating with the star can be represented as the sum of the gravitational and centrifugal forces,}
\begin{equation}
{\Phi(r)=\Phi_{\rm g}+\Phi_{\rm centri}=-\frac{GM_*}{r-\rg}-\frac{\Omega^2 r^2 {\rm{sin}}^2\theta}{2}.}
\label{eq:effectGrav}
\end{equation}
{On simplifying the above equation using the definition of $\rd$ from Eq.~(\ref{eq:rd}) and the relations $r_{\rm co} \equiv \left(GM_*/\Omega^{2}\right)^{1/3}$ and $\alpha \equiv r_{\rm co}/\rd$, we get,}
\begin{equation}
{\Phi(r)=-\Omega^2 r_{\rm co}^2 \left [\frac{\alpha \rd}{r-\rg}+\frac{(r/\rd)^3}{2\alpha^2} \right].}
\label{eq:phi_sim}
\end{equation}
{Now imposing the strong magnetic field condition (Eqs. \ref{eq:ombfi3}) and using Eq.~(\ref{eq:phi_sim}), we simplify the generalised Bernoulli constant defined along a specific stream line $\Psi(r,\theta)=$ constant, given in Eq.~(\ref{eq:bernoulli}) to,}
\begin{equation}
{E(\Psi)=\frac{1}{2}\vp^{2}+h+\Phi(r)+\int\frac{\Delta Q }{\rho \vp}{ds}.}
\label{eq:brenoulli_sim}
\end{equation}
{In the above equation and in the rest of the paper, we have considered $\alpha=1$.
This expression is a constant of motion even in case of a dissipative flow.
If dissipation is absent ($\Delta Q=0$), the above equation reduces to what is called the canonical form of Bernoulli constant and is given by,}
\begin{equation}
{{\cal E}=\frac{1}{2}\vp^{2}+h+\Phi(r).}
\label{eq:brenoulli_cano}
\end{equation}

\subsection{Relativistic EoS and the form of thermodynamic variables}
We need an EoS which relates the thermodynamic variables. 
As discussed before, we used the relativistic EoS for multiple species flow with variable adiabatic indices given by 
\citeauthor{cr09} (CR) in 2009. The form is given as follows,
\begin{equation}
{e} = \sum_{i}e_{i} = \Sigma_{i}\left[n_{i}m_{i}c^{2} 
+ p_{i}\left(\frac{9p_{i} + 3n_{i}m_{i}c^{2}}{3p_{i} + 2n_{i}m_{i}c^{2}}\right)\right],
\label{eq:eos}
\end{equation}
where, $n$ is the number density and index `$i$' suggests a sum over the species that constitute the plasma. {We have considered in our work a fully ionised plasma. Since, hydrogen is the most abundant element in the universe, an ionised flow is composed of electrons (e) and protons (p). We ignore in the present work the presence of pair production and its corresponding annihilation. Thus, the presence of positrons inside the system is negligible.}
We assume the plasma to be neutral, therefore
$\ne=\np$. On simplifying Eq.~(\ref{eq:eos}) using the expressions for $n$, $\rho$ and $p$ (\citetalias{scp20}) we get,
\begin{equation}
{e} = \rhoe c^2 \left( f_{\rm e} + \frac{f_{\rm p}}{\eta}\right)=\frac{ \rho c^2f}{\tilde{K}},
\label{eq:eos2}
\end{equation}
where,  $f_{i}=1+\Theta_i\left(\frac{9\Theta_i+3}{3\Theta_i+2}\right)$ and $f=f_{\rm e}+f_{\rm p}/\eta$. Here, $\eta = \me/\mpr$ and $\tilde{K}=1+1/\eta$. $\thetae=k\te/(\me c^2)$ and $\thetap=k\tp/(\mpr c^2)$ are dimensionless 
temperature of electron and proton respectively.\\
The adiabatic index for each species is self-consistently calculated from their temperature, using the following equation,
\begin{equation}
\Gamma_i = 1 + \frac{1}{N_i},
\label{eq:adi}
\end{equation}
where,  $N_i = {d f_i}/{d \Theta_i}$ is the polytropic index.

 
\subsection{Final form of the equations of motion for two temperature flow}
\label{sec:eom} 

In the equations to follow, we drop all the subscripts `po' which stands for poloidal components and represent the variables $\vp$ with $v$, $\bp$ with $B$, $\ap$ with $A$ and so on. \\
The differential equations for electron and proton temperatures are obtained by simplifying the first law of thermodynamics (Eq.~\ref{eq:flt}) using Eq. \ref{eq:ds} and  the EoS (Eq. \ref{eq:eos}) to get: 
\begin{eqnarray}
\frac{d \thetae}{dr} &=&-\frac{\thetae}{N_{\rm e}}\left[ \frac{1}{v}\frac{d v}{dr} +\frac{3}{2r}
\left(\frac{8-5r/\rd}{4-3r/\rd}\right)\right] - {\ebb},
\label{eq:dtedr}
\\
\frac{d \thetap}{dr} &=&-\frac{\thetap}{N_{\rm p}}\left[ \frac{1}{v}\frac{d v}{dr} +\frac{3}{2r}
\left(\frac{8-5r/\rd}{4-3r/\rd}\right)\right] - \eta{{\pbb}},
\label{eq:dtdr}
\end{eqnarray}
where, ${\ebb =\Delta Q_{\rm e}\tilde{K}/(\rho v{N_{\rm e}})~(ds/dr)}$ and ${\pbb =\Delta Q_{\rm p}\tilde{K}/(\rho v{N_{\rm p}})~(ds/dr)}$. The $Q_i$s used here are in dimensionless form which is derived from their dimensional counterparts $\bar{Q}_i$ (units of erg cm$^{-3}$ s$^{-1}$) using the relation $\bar{Q}_i\rs/(\bar{\rho} c^3)$ where ${\bar{\rho}}$ is the mass density of the species in units of g/cm$^3$.} 

Simplifying Euler equation (Eq.~\ref{eq:euler}) {and substituting Eqs. \ref{eq:accrate}--\ref{eq:constk}, \ref{eq:rd}, \ref{eq:bp}, \ref{eq:phi_sim}, \ref{eq:eos2}--\ref{eq:dtdr}, }
gives the {gradient of} poloidal velocity,
\begin{equation}
\frac{d v}{dr}=
\frac{{\cal N}(r,v,\thetae,\thetap)}{{\cal D}(r,v,\thetae,\thetap)},
\label{eq:dvdr}
\end{equation}
where,
\begin{eqnarray}
{\cal N} &=& \frac{3\as^2}{2r}\left( \frac{8-5r/\rd}{4-3r/\rd}\right) + \frac{\ebb+\pbb}{\tilde{K}} - \Phi^{'}\nonumber \\ 
{\cal D}&=&v^2\left(1 - \frac{\as^2}{v^2} \right).\nonumber
\label{eq:numden}
\end{eqnarray}
We have defined the speed of sound in two-temperature magnetized flow as, $\as^2=(\Gamma_{\rm e} \thetae + {\Gamma_{\rm p} \thetap}/{\eta})/\tilde{K}$.

\subsection{Emission processes}
\label{sec:rad}
In this section, we discuss the radiative processes that are mainly responsible for the heating
and cooling of protons and electrons present in the flow. 
Coulomb coupling is the main mechanism responsible for energy exchange between protons and electrons. It generally serves as a cooling term for protons and heating term for electrons. Therefore, $Q^-_{\rm p} = Q^+_{\rm e} = Q_{\rm cc}$. We assumed bremsstrahlung ($\qbr$) and synchrotron ($\qsyn$) as the main emission processes responsible for the cooling of electrons. 
The soft photons generated from these processes may upscatter to higher energies on interacting with 
energetic electrons through a process called inverse-Comptonization ($\qcbr$ and $\qcsy$). This radiative process leads to further cooling of electrons. 
Thus, $\qem=\qbr+\qsyn+\qcsy+\qcbr$. The seed photons  generated by   bremsstrahlung and synchrotron process can also heat up the electrons through a process called Compton heating \citep{esin97}. This happens when the energy of the seed photons is greater than the thermal energy of the electrons present in the flow. 
This term  ($Q_{\rm comp}$) 
serves as a heating term for electrons rather than cooling. The expressions of all the above radiative processes have been given in \citetalias{sc19a,sc19b,scp20,sc22} and references therein.
\subsubsection*{Black body emission  and its Comptonization}
Accreted matter on settling down onto the star's poles can form a thermal mound. This optically thick mound is a source of  blackbody photons {which serves as an additional spectral component}. These soft photons on encountering electrons present in the post-shock accretion flow can get Comptonized. 
The formula used to calculate the height and temperature of the thermal mound is from \cite{bw07}. We follow their prescription to obtain the corresponding emissivity and spectrum.

\subsection{Spectral analysis}
\label{sec:spec}

{The methodology followed to obtain the spectrum as seen by a distant observer is similar to that  as described in \cite{scp20,sc22} (also, see \cite{shap73}). We briefly discuss it here. First we compute the emission, which is the isotropic emissivity per unit frequency per unit solid angle,  in the fluid rest frame.  This is represented by} $j_\nu(r)$ {and is computed at each radius of the flow. The unit is ergs s$^{-1}$ cm$^{-3}$ sterad$^{-1}$ Hz$^{-1}$. The emissivity includes contribution from the different dissipative processes present inside the flow. Thus, $j_\nu =j_{\nu|{\rm br}}+j_{\nu|{\rm sy}}+j_{\nu|{\rm cbr}}+j_{\nu|{\rm csy}}+j_{\nu|{\rm bb}}+j_{\nu|{\rm cbb}}$. We use special-relativistic transformations to convert this $j_\nu$ from fluid rest frame to a local flat frame ($j{'}_{\nu '}$). The expressions for this transformation are given by: ${j{'}_{\nu '}=j_\nu \frac{1-v^2}{(1-v {\cos} \theta ')^2}~~~~~\mbox{and}~~~~ \nu{'}=\nu \frac{\sqrt{1-v^2}}{(1-v {\cos} \theta ')}}$. Here, $\theta '$ is the angle between the flow velocity ($v$) directed inwards towards the central object and the line of sight. On integrating the above expression for $j{'}_{\nu '}$ over the whole volume of the flux tube and on all solid angles we get the luminosity ($L_\nu$) of the system per unit frequency interval. We have also included the effect of gravitational redshift which introduces a factor of $\sqrt{1-2/r}$ in the observed frequency. On integrating $L_\nu$ over all frequencies we get the bolometric luminosity. In this work we have presented the spectra in terms of $L_\nu/(h\nu)$ (in units of keV$^{-1}$ s$^{-1}$) vs $h\nu$ (keV) for a better representation. We note here that the expressions of $j_\nu$ for bremsstrahlung, synchrotron and their respective Comptonizations have been taken from \cite{rl86,man97,wz00}. For computation of the emissivity of blackbody (BB) radiation from the thermal mound and its Comptonization we use the prescription followed by \cite{bw07} and \cite{ss75}. The amount of  BB radiation depends on the height and width of the mound formed on the surface of the NS, the expressions of which are adopted from \cite{bw07}. 
}

\subsection{Entropy accretion rate expression}
\label{sec:entropy}
Here we discuss and derive  the entropy accretion rate formula for two-temperature accretion flows around magnetized stars. The derivation is exactly same as was in case of BHs \citepalias{sc19a,scp20}. Let us assume adiabaticity of protons and electrons and  remove all the explicit heating and cooling terms  present in the first law of thermodynamics (Eq. \ref{eq:flt}), which as discussed before is defined separately for electrons and protons in the two-temperature theory. Thus, we have,
\begin{align}
\frac{d\thetap}{dr}=\frac{\thetap}{\polyp}\frac{1}{\np}\frac{d\np}{dr} + \frac{\qcc \eta \tilde{K}}{\rho v \polyp} \nonumber \\ 
{\rm{and}}~~
 \frac{d\thetae}{dr}=\frac{\thetae}{\polye}\frac{1}{\nel}\frac{d\nel}{dr} -
 \frac{\qcc\tilde{K}}{\rho v \polye}.
\label{eq:fst1law2}
\end{align}
This equation cannot be integrated analytically, due to the presence of Coulomb coupling term, unlike in case of one-temperature flows, where $\tp=\te\Rightarrow \qcc=0$ and we have an analytical expression for entropy \citepalias{sc18a,sc18b}.

However, it is possible to integrate Eq. (\ref{eq:fst1law2}) where $\qcc$ is $0$ or negligible. This conjecture can only be fulfilled in regions where gravity dominates any other process or interaction. The strong gravity implies that infall timescales in these regions would be shorter than any other timescales, such that before any dissipation processes act, the matter would be advected towards the central object. It is also in this region that adiabatic conditions are valid. let at a distance $\rin$ the flow approaches adiabatic condition. Thus, an analytical expression of entropy is obtained by integrating Eq. (\ref{eq:fst1law2}), using the adiabaticity condition and $\qcc=0$ at $\rin$ and is given by,
\begin{align}
&\neh=\tilde{\kappa}_1 ~ {\rm exp}{\left({\frac{\feh-1}{\theh}}\right)}\theh^{\frac{3}{2}}(3\theh+2)^{\frac{3}{2}}\\
&\nph=\tilde{\kappa}_2 ~{\rm exp}{\left({\frac{\fph-1}{\thph}}\right)}\thph^{\frac{3}{2}}(3\thph+2)^{\frac{3}{2}},
\label{eq:ent_2}
\end{align}
where, $\tilde{\kappa}_1$ and $\tilde{\kappa}_2$ are constants which measure entropy. Subscript `${\rm in}$' defines quantities at $\rin$ where the above assumptions hold true.\\
Charge neutrality of the accretion flow suggests $\neh=\nph=\nh$.  Therefore, we can write,
\begin{equation}
\nh^2 =\neh \nph \Rightarrow \nh =\sqrt{\neh \nph}
\end{equation}
Thus, 
\begin{align}
\nh=\tilde{\kappa} \sqrt{{\rm exp}{\left({\frac{\feh-1}{\theh}}\right)}~{\rm exp}{\left({\frac{\fph-1}{\thph}}\right)}\theh^{\frac{3}{2}}\thph^{\frac{3}{2}}{(3\theh+2)^{\frac{3}{2}}}  (3\thph+2)^{\frac{3}{2}}},
\end{align}
where, $\tilde{\kappa}=\sqrt{\tilde{\kappa}_1 \tilde{\kappa}_2}$.\\
Thus, the expression for entropy accretion rate, using Eq.~(\ref{eq:accrate}), can be written as,
\begin{align}
\nonumber \mdotin &=\frac{\dot{M}}{\tilde{\kappa} (\me+\mpr)}\\ 
&=\sqrt{{\rm exp}{\left({\frac{\feh-1}{\theh}}\right)}{\rm exp}{\left({\frac{\fph-1}{\thph}}\right)}\theh^{\frac{3}{2}}\thph^{\frac{3}{2}}{(3\theh+2)^{\frac{3}{2}}}} \nonumber\\
& \times \sqrt{{(3\thph+2)^{\frac{3}{2}}}}  v_{\rm in}A_{\rm in}.
\label{eq:ear}
\end{align}
This formula is exactly similar to what was obtained for BH accretion flows \citepalias{sc19a,scp20}.
The importance and use of this formula in two-temperature flows around magnetized stars have been discussed briefly in section \ref{sec:meth}.

\subsection{Sonic point conditions and shock conditions}
\label{sec:spshock}

Accretion flow  around magnetized stars are generally transonic in nature, similar to flows around BHs  (\citealp{kol02,ka08}; \citetalias{sc18a,sc18b}). 
The flow starts with a subsonic velocity, $v < a_{\rm s}$ at  $r\approx\rco$.
As the matter gets accreted along the magnetic field lines 
the increase in
gravitational potential energy of the matter leads to a corresponding increase in its kinetic energy. This increases the flow velocity. Additionally matter is also compressed to smaller and smaller volume as it gets accreted. Thus, as a secondary effect, the
temperature and hence the sound speed increases. At a certain point of the flow $\rc$, the flow velocity is equal to the local sound speed
$ v_c=a_{\rm sc} \implies$ or the Mach number at $\rc$,
$M_c=1$, where $M_c=v_c/a_{\rm sc}$. This point is called the critical point of the flow. From
Eq. (\ref{eq:dvdr}), we see that when $v_c=a_{\rm sc}$ $ \implies {\cal D}=0$. For the flow to be smooth and continuous, 
${\cal N}$ should also go to $0$. Thus, at $\rc$ the velocity slope has the form, $dv/dr\rightarrow 0/0$. 
Now, because of the  rotation of  the star, a centrifugal force acts in opposite direction to gravity. 
This induces the formation of multiple critical (sonic) points (MCPs). 
The critical points are named according to the distance from the central 
object: inner ($\rci$), middle ($\rcm$) and outer ($\rcou$). Out of these, $\rci$ and $\rcou$ are X-type critical points and are physical in nature, while $\rcm$  is unphysical (spiral-type or O-type depending on whether the system is dissipative or not, respectively) and matter cannot flow through it. For X-type critical 
points, 
$dM/dr|_{\rc}$ possess two real roots: one $<0$, which corresponds to the accretion solution 
and the other $>0$, which is called the excretion solution.  In the MCP regime an accretion flow can also harbour shocks. These shocks are called secondary shocks ($\rss$) and are formed for certain combination  of flow parameters, driven by the centrifugal and pressure
gradient forces.  These have been reported recently in the single temperature
regime by \citetalias{sc18a,sc18b}.

Apart from strong, ordered magnetic field of
a magnetized star, the presence of a hard surface is an additional  major difference, that distinguishes it from a BH.
The accretion flow although accelerated to achieve transonicity by its strong gravity, has to settle down
onto the star's surface. This drives a terminating shock, also known as primary shock ($\rps$) (also  see, Fig. \ref{fig:1}, dark maroon region). These are formed very close to the star's surface while 
secondary shocks are formed anywhere  between the  primary shock and the
co-rotation radius (i.e., $\rps<\rss<\rco$).
A primary shock is always formed and forces the flow to satisfy the star's surface boundary condition. However, secondary shocks are formed for a particular set of flow parameters.

In the strong magnetic field regime the MHD shock conditions \citep{ke89} reduce to hydrodynamic shock conditions \citepalias{sc18a, sc18b}
and  are given by,
\begin{eqnarray}
{\rm Conservation~of~mass~flux~:~}&\left[\rho v\right]&=0,\label{eq:shockcond1} \\
{\rm Conservation~of~momentum~flux~:~}&\left[\rho v^{2} + p\right]&=0,\label{eq:shockcond2}\\
{\rm Conservation~of~energy~flux~:~}&\left[{\dot{E}} \right]&=0.
\label{eq:shockcond3}
\end{eqnarray}
where the square brackets imply the difference between the pre-shock and post-shock flow variables.

\section{Methodology}
\label{sec:meth}

Here, we present first, the methodology to find general transonic two-temperature accretion solutions around strongly magnetized stars and then elaborately examine the {problem} of degeneracy present in {two-temperature} theory. After {that}, we  discuss the use of entropy accretion rate formula given in section \ref{sec:entropy} to remove the degeneracy present in these type of solutions.

\subsection{Methodology to obtain general two-temperature transonic solution around a strongly magnetized star}
\label{sec:meth2}

In this section we discuss the methodology to obtain general two-temperature accretion solutions {onto a magnetised compact star, for a}
given set of CoM. But before going into the details it is important to remember that for a given set of CoM we will obtain multiple transonic solutions unlike in case of one-temperature flows where a unique solution is obtained \citepalias{sc18a, sc18b}. The additional temperature variable in two-temperature regime and the absence of any equation relating this temperature with the other flow variables is responsible for {the} multiplicity of solutions. The entropy maximisation formulation which will be discussed in the upcoming section will serve as a tool to select only one solution out of all. Since the system we are working on is dissipative, we cannot have a measure of entropy and the entropy accretion rate form (Eq. \ref{eq:ear}) can only be used at a point $\rin$ where the gravity is very strong and the matter velocities achieve free-fall velocity. Therefore, keeping in mind the importance of the point $\rin$ we discuss here the methodology to find general solutions.




\subsubsection{Utilising the property of gravity to find $\rin$ $\rightarrow$ the inner boundary}
\label{sec:methgrav}
The gravitational pull on any particle of unit mass by an object of mass $M_*$, depends on the value of $M_*$ and also on the distance between
the centre of gravities of these two masses. 
Utilising this property we can conclude that, if a star of mass $M_*$ and radius $r_*$, is confined in 
a radius $\rin~(<r_*)$, then the gravitational force experienced {by a mass} at a point $r$ (where, $r>r_*>\rin$) would be exactly the same, irrespective of whether the radius of the star is  $r_*$ or $\rin$. 

{We find that it is at  $\rin~\rightarrow ~\rg$ that the infalling matter will asymptotically achieve free fall velocity and nowhere else. 
We select $\rin$ as the inner boundary while obtaining solutions, assuming the star's mass to be concentrated within $\rin$. Once a  solution is obtained, the portion of the  solution within  $\rin~\leq~r\leq~r_*$ is called the ghost solution. We check for the terminating shock of the transonic solution in  regions $r> r_*$.
}

{In order to check whether this methodology to obtain solution works correctly, we obtain solutions in one temperature regime using this method (as described in section 3.1.2), and compare it with the solution obtained by \cite{sc18b}, who used sonic point analysis method (see Appendix \ref{app:1}).
From the comparison of solutions in Fig. \ref{fig:a1}, it is clear that both the methods are equivalent.} 

\subsubsection{Steps to find a general TS or transonic solution}
\label{sec:steps}
\begin{enumerate}
\item Supply the CoM ($E$, $P$, $\mdot$). Also, we need to supply surface magnetic field $B_*$ of the magnetized star of mass $M_*$ and radius $r_*$. 
\item {We consider an inner boundary point $\rin \sim \rg<r_*$.} 
\item We fix a value of $\thph$ {at $\rin$}. We note here that this is an extra variable present in two-temperature flows and was not present in one-temperature systems. Hence, before starting the methodology to find a global transonic solution (TS), we have to first fix a particular value of $\thph$.
\item We supply an initial guess value of $\theh$ at $\rin$.
\item We obtain the exact value of $\vin$, from the Bernoulli parameter expression obtained by equating of $E\equiv{\cal{E}}(\vin,\thph,\theh)$ (Eq.~\ref{eq:brenoulli_cano}). 
\item With the initial values of $\vin,\thph,\theh$ at $\rin$, we evaluate $dv/dr$, $d\thetap/dr$ and $d\thetae/dr$
(Eqs.~\ref{eq:dtedr}-\ref{eq:dvdr}) and then integrate these equations from $r=\rin$
outwards, with increasing $r$. 
\item There is a high probability that the guess value of $\theh$ might lead to a completely supersonic branch (SB) solution or a
multi-valued branch (MVB) solution (similar to  dotted blue curve or dashed dotted green of Fig.~\ref{fig:a1}b in
one-temperature case).
So we iterate $\theh$ until the solution passes through a sonic
point ($\rc$). 
\item {Once $\rc$ is found, 
$dv/dr|_{\rc}$ is obtained by employing L'Hospital's rule. 
Thereafter,  we further integrate outwards till the co-rotation radius ($\rco$) is reached. In this way, we obtain the full global TS. }
\item It may be noted that there might be MCP for the same set of CoM and $\thph$.
For the same value of $\thph$, which has produced a TS, we continue to search for other sonic points by changing the guess value of $\theh$ by a large factor and then repeating
steps (iv)--(viii).  

\item {Once the TS is found, it needs to satisfy the star's surface boundary conditions}. 
For every $r>r_*$ we impose a shock jump and using the corresponding subsonic post-shock values of the flow variables (obtained using Eqs. \ref{eq:shockcond1}--\ref{eq:shockcond3}), we integrate the EoM inwards towards the star's surface until $r\rightarrow r_*$ is reached.  We select the $r$ for which the {flow} velocity at $r_*$ reaches negligible values or the surface boundary conditions are satisfied \citep{dhang21}.  
This is the primary shock location ($\rps$).  
\end{enumerate}

\begin{figure}
\centering
\hspace{0.0cm}
\includegraphics[width=9.cm,trim={0.1cm 0.6cm 0.cm 2.2cm},clip]{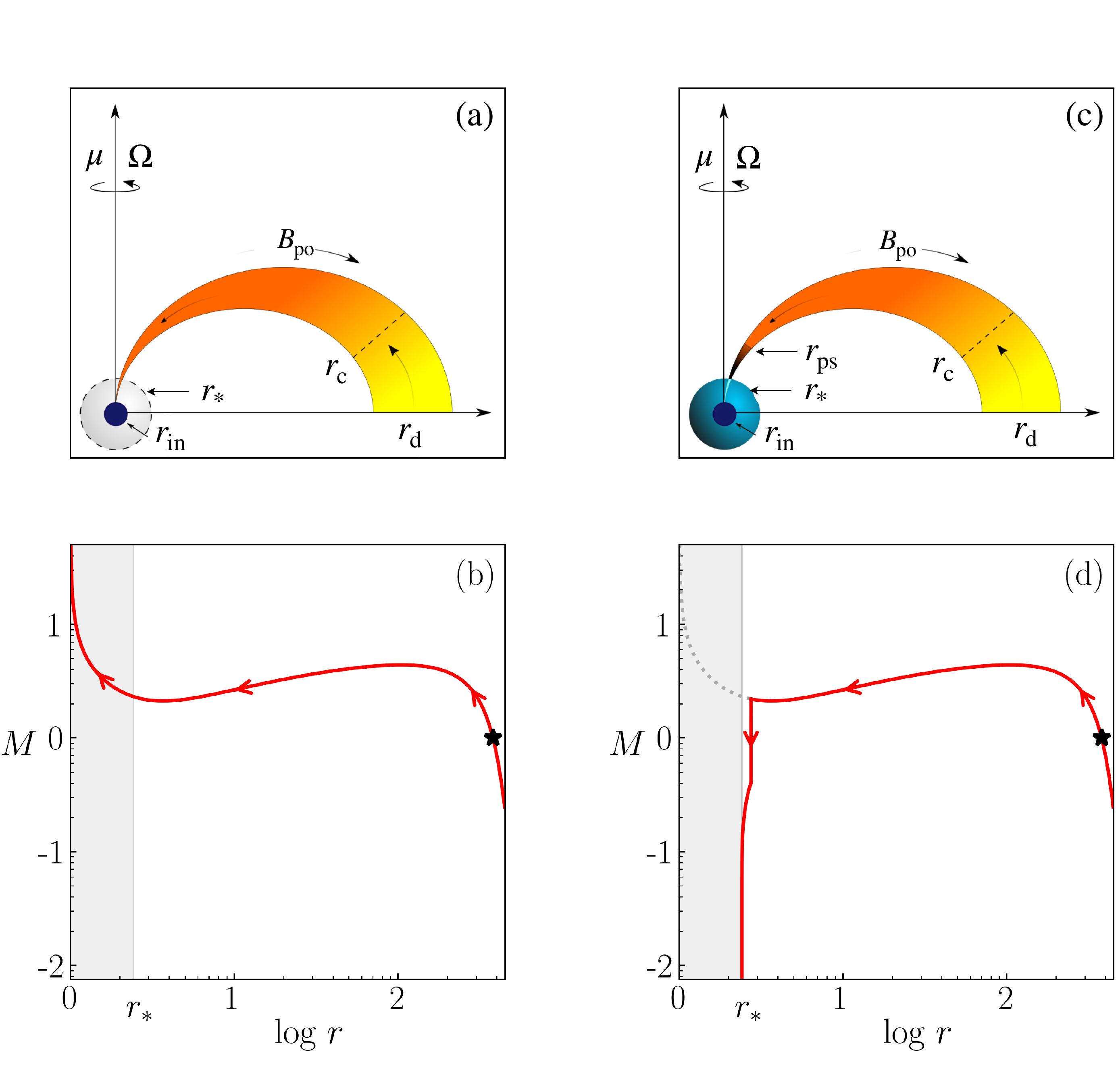}
\vspace{0.0cm}
\caption[] {\small Panel (a) and (c) shows an accretion flow geometry where matter is accreted until it reaches a  radius, $\rin\sim \rg$ ($<r_*$) and $r_*$ respectively. In both cases the central object has a mass $M_*$.
Panel (b) plots the TS solution corresponding to panel (a).  
When this flow satisfies the star's surface boundary condition, the solution undergoes a primary shock transition near the surface  and is plotted in panel (d). The `ghost solution' is represented in dotted grey. The flow parameters are given in text below. 
}
\label{fig:2}
\end{figure}

We note that throughout the aforementioned steps used for obtaining the TS, we have kept $\thph$ fixed. Thus, the global TS obtained can be identified using the value of $\thph$ apart from the supplied CoM.
The above methodology is illustrated more elaborately in Figs.~\ref{fig:2}a--d. In Figs.~\ref{fig:2}a and c, cartoon diagram of an accretion flow around a magnetized star is presented. 
In Fig.~\ref{fig:2}a, the methodology to  find the projected solution or the ghost solution is illustrated.
The star surface $r_*$
is presented in dashed circle and $\rin$ is the supposed radius which contains the same mass $=M_*$ of the star but in a {smaller volume. We consider $\rin \sim \rg$,
so that $v \rightarrow$ free-fall.} .
{{Figure}.~\ref{fig:2}b plots the TS obtained following steps (i)--(ix), with  the sonic point $\rc$ represented by a black star.}  
The flow parameters used for obtaining the solution 
are, $E=0.9985$, $P=1.16$s, $\mdot=10^{15}$g/s, $B_*=10^{10}$G,  $M_*=1.4\msol$ and $r_*=10^6$cm. The  value of proton temperature at $\rin$ for this solution  is $\tph=1\times 10^{11}$K ($=\thph \mpr c^2/k$).  
Once the complete TS is obtained, one may  find the location of the primary shock ($\rps$), which is generally formed close to 
$r_*$. We represent the situation via a cartoon diagram in Fig.~\ref{fig:2}c, where after obtaining the full TS, the shock {conditions} were satisfied at $\rps=3.301\rg$ (dark maroon shade being the post-shock flow), see {step (x)}. 

In this way, we obtain a global two-temperature accretion solution around a magnetized star  satisfying surface boundary conditions 
(see Fig.~\ref{fig:2}d). %

\subsection{Finding the unique two-temperature accretion solution: Problem of degeneracy and the methodology to constrain it}
\label{sec:meth1}

{As have been discussed before, an increase in variable (an additional temperature) in two-temperature regime is not compensated with any increase in the number of equations or any 
additional relation, which relates it with the other flow variables at any boundary. The presence of an extra temperature variable is responsible for {the} degeneracy. This problem is generic in two-temperature theory and does not depend on the type of central object considered.} {We saw in step (iii)  of the above section that $\thph$ was fixed for obtaining a particular global accretion solution, apart from the specified set of  CoM.} Keeping the CoM same, if $\thph$ is varied we would get a completely different global solution. This suggests that  for a given set of CoM, we would get multiple accretion solutions each with different $\rc$ and their properties. It can hence be concluded that two-temperature flows are degenerate in nature.

\subsubsection{Entropy maximisation methodology in case of BHs}
The problem was investigated in details by \citetalias{sc19a,sc19b,scp20,sc22} for the case of two-temperature flows around BHs. Here, the degeneracy was constrained using the first principles, without taking recourse to any arbitrary assumption.}
In the absence of any physical principle constraining the relation between the temperatures, \citetalias{sc19a,sc19b,scp20} utilised the concept of entropy to obtain a unique transonic two-temperature accretion solution.
However, because of the presence of electron -- proton energy exchange term or the Coulomb coupling term in the first law of
thermodynamics, one cannot obtain an analytical expression for entropy measure. But it is known that near the BH horizon,
strong gravity overwhelms any other interactions. Matter just outside the horizon falls freely and
the infall timescales are shorter than  cooling or Coulomb coupling time scales. Therefore, asymptotically close
to the horizon, an analytical
expression of entropy is admissible. Using this formula for entropy, strictly valid near the horizon, entropies of all the degenerate solutions were measured. 
It was found that the entropy  maximised for a certain solution.
Following the second law of thermodynamics, that nature would prefer a solution 
with maximum entropy, degeneracy was removed in two-temperature accretion flows around BHs. Additionally, it was shown that the maximum entropy solution is the most stable one \citepalias{scp20}. 
\subsubsection{Entropy maximisation methodology in case of stars with hard surface}

\begin{figure}
\centering
\hspace{0.0cm}
\includegraphics[width=8.5cm,trim={0.8cm 16.6cm 11.cm 0.4cm},clip]{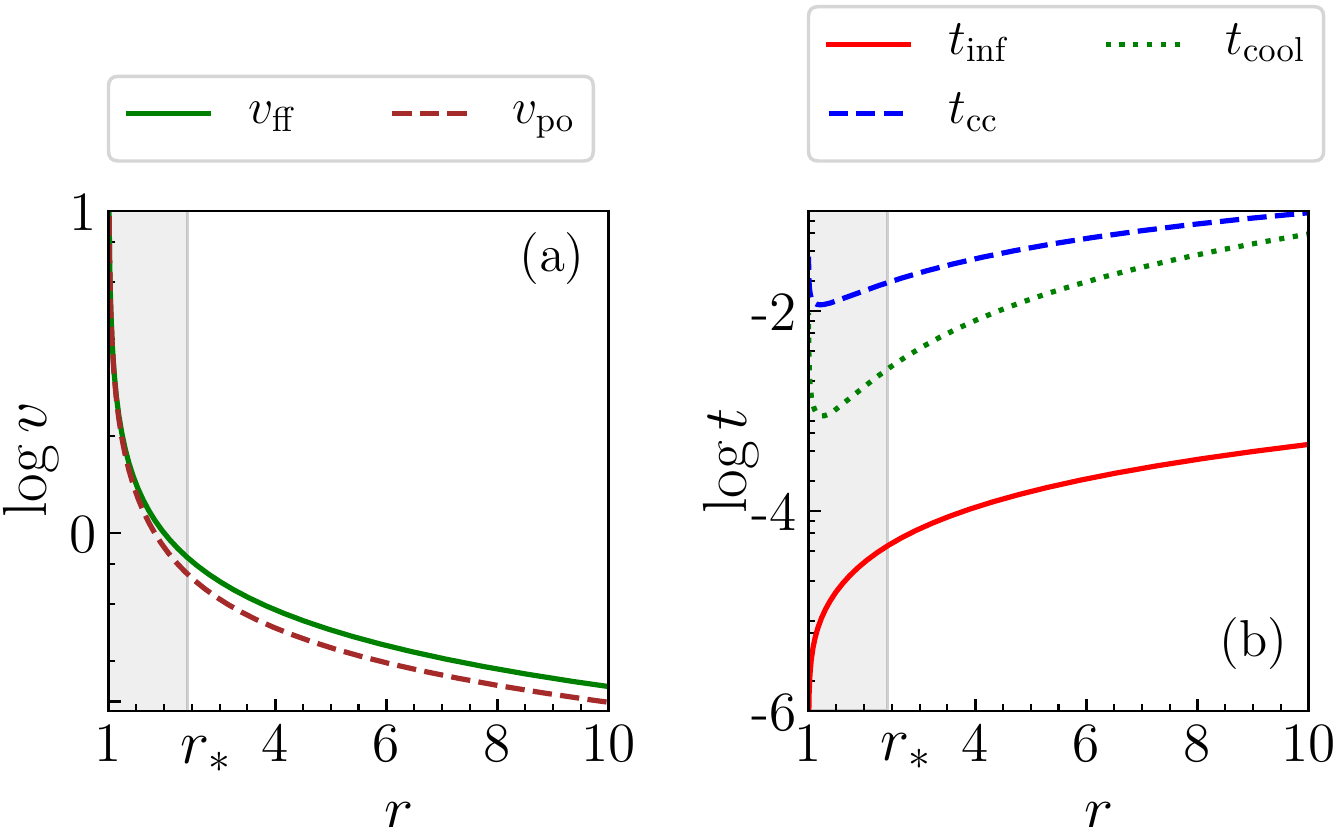} 
\vspace{0.0cm}
\caption[] {\small Comparison of (a) poloidal velocity ($\vp$) with free-fall velocity ($v_{\rm ff}$) and (b) infall timescale ($t_{\rm inf}$) with cooling ($t_{\rm cool}$) and Coulomb coupling ($t_{\rm cc}$) timescales.}
\label{fig:freefal}
\end{figure}

An inner boundary condition similar to BH is not present outside star's surface. As discussed before, strong gravity of the magnetized star can
make an accretion flow transonic, with matter approaching the surface supersonically,  however, the poloidal velocity
of the supersonic branch
is not high enough to approach free-fall values. {To illustrate, we compare the poloidal velocity $\vp$ (dashed brown) with the freefall velocity $v_{\rm ff}$ (solid green) in Fig. \ref{fig:freefal}a
for flow parameters $E=0.9983$, $P=1$s, $\mdot = 10^{15}{\rm g/s}$, $M_*=1.4\msol$, $r_*=10^6$cm and $B_* = 10^{10}$G. 
Certainly $v_{\rm ff}>\vp$ everywhere, for $r>r_*$ (except at $r\rightarrow \rin$ where $\vp\rightarrow v_{\rm ff}$, also see the discussion in Section \ref{sec:methgrav}). This is not very surprising, if we refer to Eq. \ref{eq:bernoulli}. Free fall velocity is achieved when the first term on r. h. s (infall kinetic energy) is equal to gravitational energy (fourth term in r. h. s) and all other terms are negligible.
Since other terms are present in the expression of $E$, then $\vp << v_{\rm ff}$, except at $r\rightarrow \rin$, where gravity over powers all other interactions. 
} 
Moreover, the presence of a hard surface intervenes 
the supersonic
flow and drives a shock at $\rps$ near the surface and the accreting matter finally settles down with negligible velocities.
In other words, neither
the supersonic branch, nor the post shock flow achieves free-fall velocity. Therefore, one cannot obtain an entropy measure at the star's surface
using the entropy formula given by Eq.~\ref{eq:ear}, which is valid only in regions where infall {velocity approaches free fall, i. e., where gravity over powers all other interactions}. Thus, the methodology followed in case of BHs, cannot be adopted directly 
for flows around magnetized stars.
{
Fortunately the methodology proposed in section \ref{sec:steps} where $\rin$ is used as the inner boundary solved the problem. Therefore, all the conditions needed to apply the entropy accretion rate formula (Eq. \ref{eq:ear}) are satisfied at $\rin$.  {Additionally in Fig. \ref{fig:freefal}b, we show that infall timescale (solid red) in the funnel region is very much shorter than any other timescales (cooling or Coulomb coupling). It is the consequence of  $\vp$ 
being transonic along magnetic field line. Since infall time scale is much smaller than other relevant time scales, the accretion flow will be in the two-temperature regime.}  }
\subsubsection*{Steps to constrain the degeneracy}
The following steps needs to be implemented to constrain the degeneracy present in two-temperature accretion solutions around compact objects with a hard surface. 
\begin{enumerate}
  \setcounter{enumi}{10}
\item Once we have obtained the complete TS (following steps iii to ix) we note down the entropy measure ($\mdotin$) at $\rin$ using Eq. \ref{eq:ear}.  We identify this entropy measure to the corresponding $\thph$.
\item Now, keeping  the CoM same, we change $\thph$ and repeat steps iii--ix. We obtain another TS and compute the corresponding
$\mdotin$. We carry out this process for different $\thph$s. At the end, we will have a range of $\thph$s and their  corresponding $\mdotin$s, all for 
the same set of CoM. 
\item Following the second law of thermodynamics, a unique accretion solution is the one which has the maximum entropy. Thus, we select that solution.

\end{enumerate}

\begin{figure*}
\centering
\hspace{0.0cm}
\includegraphics[height=9.cm,trim={0.cm 8.5cm 0.cm 3.3cm},clip]{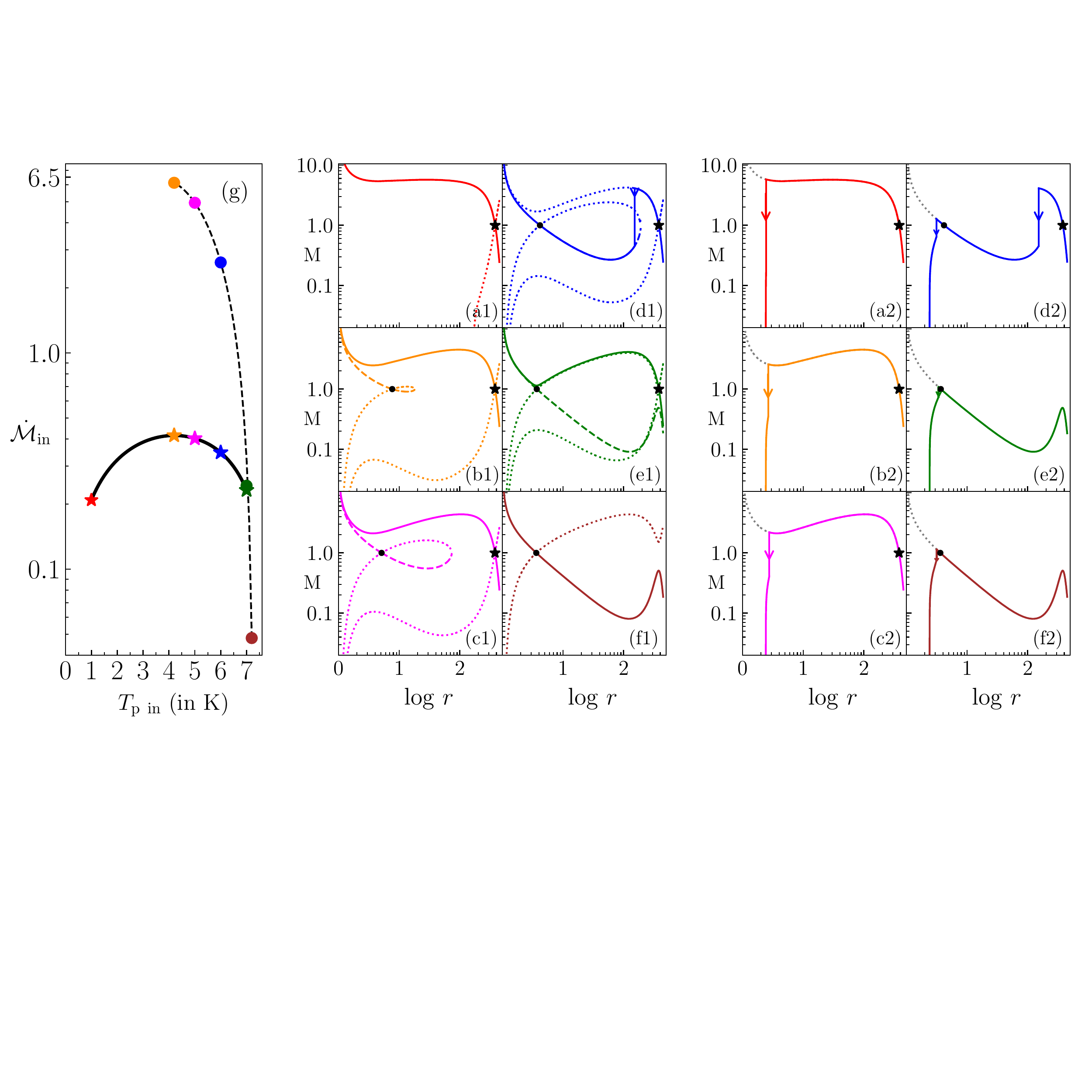}
\vspace{0.0cm}
\caption[] {\small Methodology to remove degeneracy.
Panels (a1--f1) plots the TS obtained using $\rin \sim \rg$ as inner boundary.
Panels (a2--f2) plots the same solutions, after satisfying the  surface boundary conditions at $r=r_*$.
All the solutions presented are for the same set of CoM but corresponds to different $\tph$s, values of which are $1.0\times 10^{11}K$ (red, panels a1, a2), $4.2\times 10^{11}$K (orange, panels b1, b2 ),
(c) $5.0\times 10^{11}K$ (magenta, panels c1, c2), (d) $6.0\times 10^{11}K$ (blue, panels d1, d2),
(e) $7.0\times 10^{11}K$ (green, panels e1, e2) and (f) $7.2\times 10^{11}K$ (brown, panels f1, f2).
The primary shock location (downward arrow), global (solid), excretion (dotted), non-global (dashed) and the ghost solutions
(dotted grey), respectively are indicated accordingly. 
Panel (g) plots the entropy accretion rate values $\mdotin$ for all the values of $\tph$ possible: solid black curve represents solutions passing through $\rcou$ and dashed black is for solutions passing through $\rci$. 
Entropy is maximised at $\tph=\tpm=4.2\times 10^{11}$K. CoM used are same as in Fig.~\ref{fig:2}.}
\label{fig:3}
\end{figure*}

In Figs.~\ref{fig:3}a1--f1, we plot the TSs (solid curves)  for different $\thph$s (step xii). The flow parameters used are same as in Fig.~\ref{fig:2}. It is interesting to note that all these solutions are for the same set of CoM and hence we identify each solution using the corresponding value of $\thph$. Their corresponding solutions which satisfy the surface boundary condition, are plotted in  Figs.~\ref{fig:3}a2--f2. The dotted grey curve in each of these plots represent the ghost solutions. 
Values of $\tph$ used to obtain the different TS are,
$1\times 10^{11}$K (Figs.~\ref{fig:3}a1, a2, red), $4.2\times 10^{11}$K (Figs.~\ref{fig:3}b1, b2, orange), $5.0\times 10^{11}$K
(Figs.~\ref{fig:3}c1, c2, magenta), $6.0\times 10^{11}$K (Figs.~\ref{fig:3}d1, d2, blue), $7.0\times 10^{11}$K (Figs.~\ref{fig:3}e1, e2, green),
and $7.2\times 10^{11}$K (Figs.~\ref{fig:3}f1, f2, brown). Solid curves represent the accretion solution, dotted lines represent the excretion solution which is obtained because of the presence of two roots at $dv/dr|_{\rc}$ and dashed curves represent accretion solutions which are not global. 
Accretion solutions presented in Figs.~\ref{fig:3}a1, a2 and f1, f2, possess single sonic point, where it is outer type
for the former ($\rcou$ marked with black star) and
inner type for the later ($\rci$ marked with black circle). Rest of the solutions presented in  Figs.~\ref{fig:3}b1, b2--e1, e2, have MCPs (black stars
and circles). There is also a centrifugal force driven shock transition in Figs.~\ref{fig:3}d1, d2, which is called a secondary shock ($\rss$). 
{
The global solution in this case first passes through $\rcou$, encounters a shock jump at $\rss$ and then again passes through $\rci$ and then settles down onto the surface after encountering a terminal shock at $\rps$.}

It can hence be concluded that, by just varying the inner boundary values ($\thph$ at $\rin$) we  obtain different topology of solutions, but all for the same set of CoM. %
But this should not be the case and a given set of CoM should necessarily harbour a unique solution. To remove this degeneracy we plot the entropy measure $\mdotin$ at $\rin$
vs $\tph$ in Fig.~\ref{fig:3}g (step xi). The solid black curve represent entropies of solutions having outer sonic points and  dashed is for solutions with inner sonic points. For few $\tph$s both $\rci$ and $\rcou$ exist. The coloured stars and circles correspond to $\mdotin$s of TS plotted
in Figs.~\ref{fig:3}a1, a2--f1, f2 (colour coding kept the same). 
$\mdotin$ is found to maximise for $\tph=4.2\times 10^{11}$K$=\tpm$. The TS corresponding to $\tpm$ is plotted in Fig.~\ref{fig:3}b1 and after satisfying the surface boundary conditions, which is the final  global accretion solution,  is plotted in Fig.~\ref{fig:3}b2. Although this set of flow parameters and $\thph$ harbours multiple sonic point (orange star and orange circle),  but due to the absence of a secondary shock transition the global accretion solution passes through $\rcou$ {only} (solid, orange). 

Thus, to conclude, the maximum entropy solution is the one which nature would prefer. Following the second law of thermodynamics, we constrained the degeneracy (step xiii). 
As has been shown by \citetalias{sc19a,sc19b,scp20,sc22} (but for the case of BH), among all the degenerate solutions the one with  maximum entropy should be the unique accretion solution. 


\section{Results}
\label{sec:result}
In this section we analyse in detail two temperature accretion flows around strongly magnetized compact stars in the presence of dissipative processes.
We also perform their spectral analysis. The mass and radius of the compact star is assumed to be $M_* = 1.4\msol$ and $r_* = 10^6$cm$=2.418\rg$,
respectively throughout the paper, until otherwise mentioned. We chose the compact star parameters similar to a neutron star.
All the solutions presented in the subsequent sections correspond to the maximum entropy solution and has been obtained using the methodology discussed above. 


\subsection{A typical two-temperature accretion solution around a magnetised compact star}
\label{sec:gensol}

\begin{figure}
\centering
\hspace{0.0cm}
\includegraphics[width=10.cm,trim={3.2cm 4.8cm 1.4cm 0.0cm},clip]{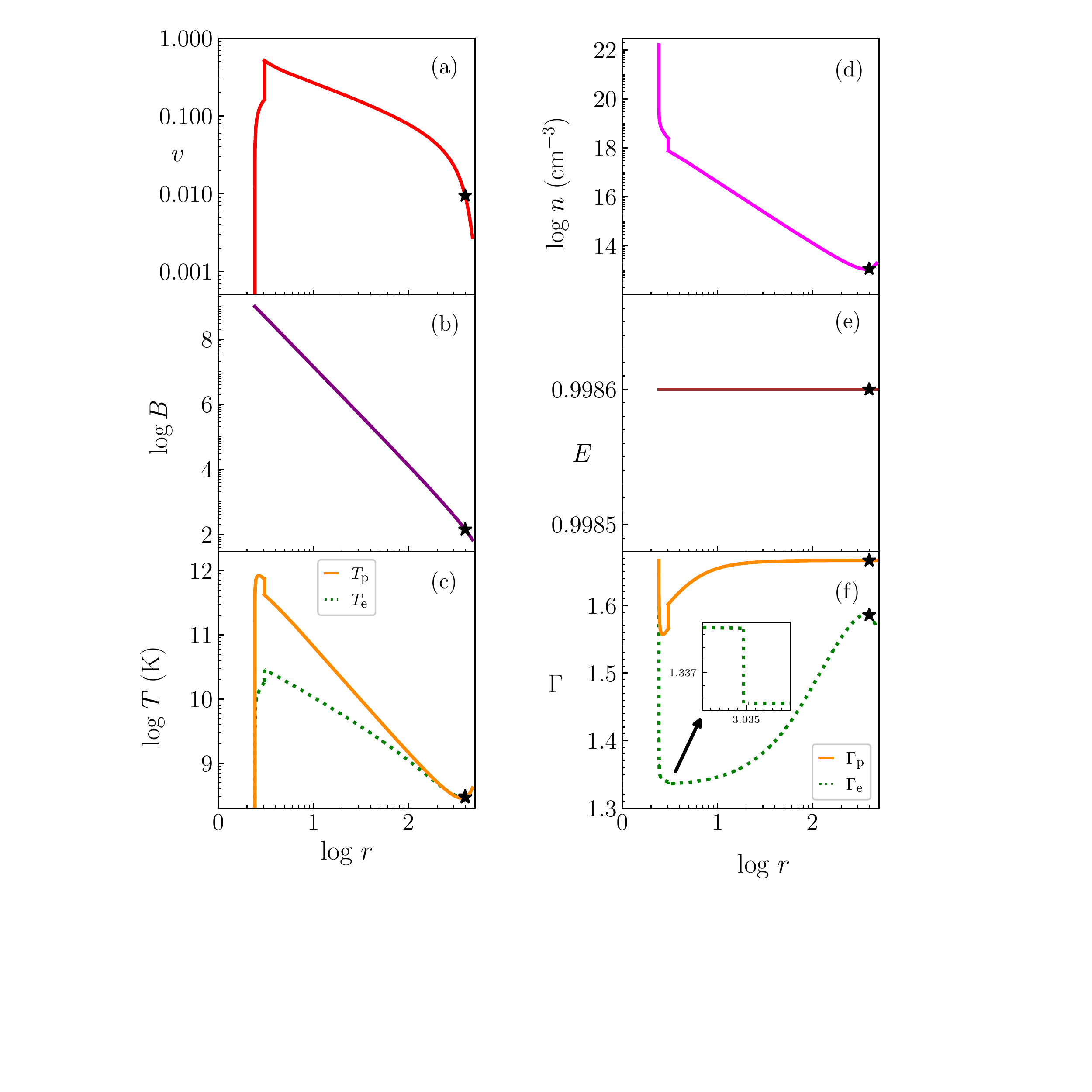} 
\vspace{0.0cm}
\caption[] {\small Flow variables corresponding to a typical two-temperature accretion solution around an compact star are plotted. 
The flow parameters are, $E = 0.9986$, $P = 1.25$s, $\mdot= 10^{14}$g/s and $B_*= 10^9$G. For these parameters there is only one sonic point ($\rcou$) marked with a black star.
}
\label{fig:4}
\end{figure}


We present a general two-temperature accretion solution in Fig.~\ref{fig:4}. 
The parameters used are, $E = 0.9986$, $P = 1.25$s, $\mdot= 10^{14}$g/s and $B_*= 10^9$G. 
We plot  flow variables in different panels which are: $v$ (Fig.~\ref{fig:4}a), $\log B$ (Fig.~\ref{fig:4}b),
$\log \tp$ and  $\log \te$ (Fig.~\ref{fig:4}c), $\log n$ (Fig.~\ref{fig:4}d), $E$ (Fig.~\ref{fig:4}e)
and $\gamp$ and $\game$ (Fig.~\ref{fig:4}f), all plotted as  function of $\log r$.
The location of  sonic point, $\rcou=393.541\rg$ is marked with a black star and 
 the location of the primary shock is at $\rps=3.035 \rg$. 
In the post-shock region, enhanced density increases the cooling processes, which decreases the temperature as well as
the velocity drastically. 
Moreover, the flow geometry (dictated by $B$, Eq.~\ref{eq:bp}) decreases faster than $r^2$, which causes the
 $n$ to increase sharply near the surface (Fig.~\ref{fig:4}d).
The temperature of the species exhibits an interesting behaviour. 
$\tp$ increases at the primary shock location due to shock heating,
while, since the electrons primarily radiate, the enhanced density at the shock front causes 
$\te$ to dip at the primary shock location (Fig.~\ref{fig:4}c). The respective adiabatic indices follows the behaviour of the
temperature distribution (Fig.~\ref{fig:4}f).  The generalized Bernoulli parameter $E$, which is a constant
of motion even in presence of dissipation is indeed found to be a constant (Fig.~\ref{fig:4}e).
\subsubsection*{Emissivities and spectrum}
\begin{figure}
\centering
\hspace{0.0cm}
\includegraphics[width=8.6cm,trim={0.8cm 11.cm 4.cm 3.2cm},clip]{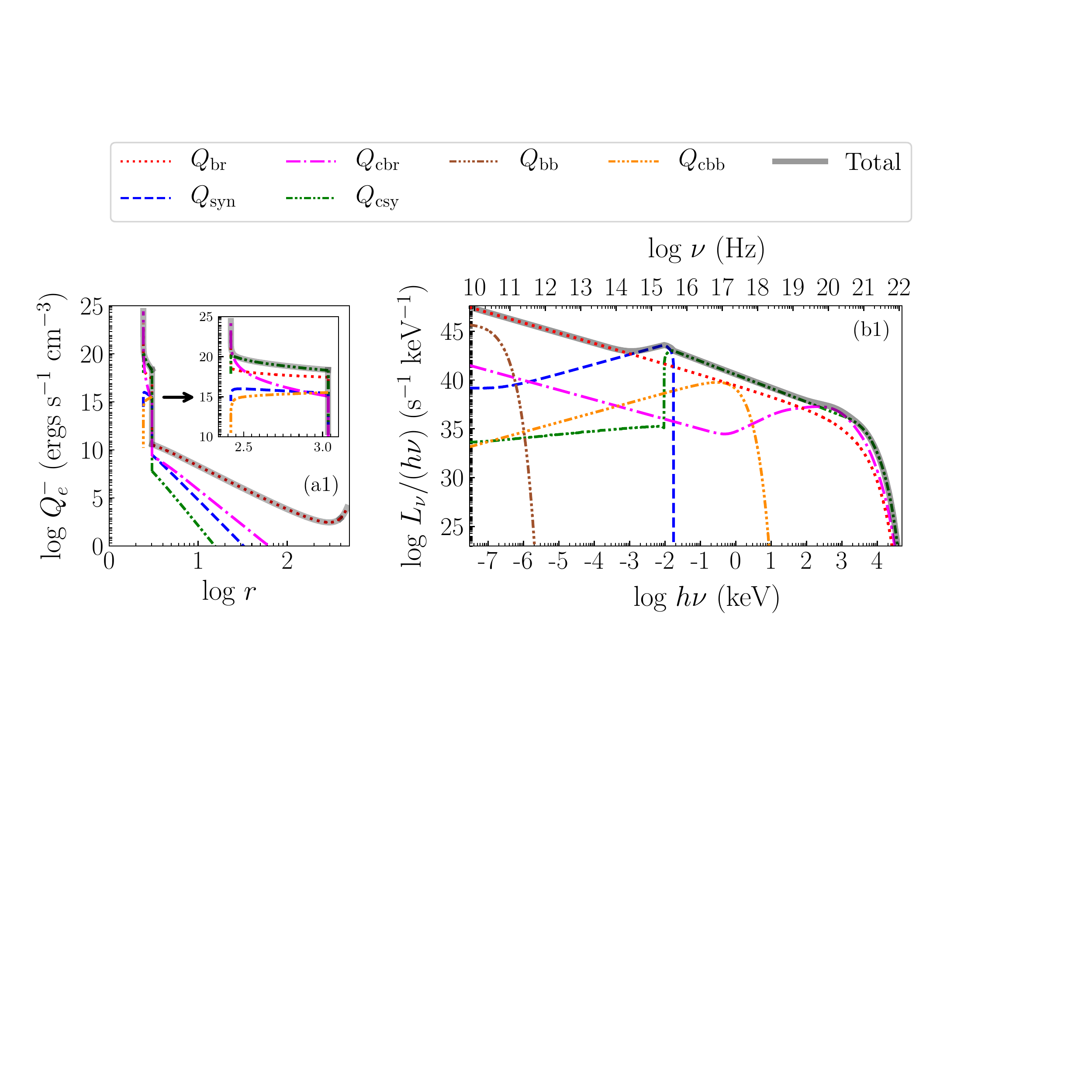}
\vspace{0.0cm}
\caption[] {\small (a) Emissivities ($\qem$) and (b) corresponding spectrum ($L_{\nu}/(h\nu)$ vs $h\nu$) are plotted.
Parameters used are same as in Fig. \ref{fig:4}.}
\vspace{-0.1cm}
\label{fig:5}
\end{figure}
In Figs. \ref{fig:5}a and b, we plot the emissivities and spectrum respectively, for different cooling mechanisms which are:
bremsstrahlung  $\qbr$ (dotted, red), synchrotron  $\qsyn$ (dashed, 
blue), Comptonized bremsstrahlung $\qcbr$ (dashed single-dotted, magenta) and Comptonized synchrotron 
$\qcsy$ (dashed double-dotted, green). Matter accreted through the magnetic funnels on settling down onto the poles of the NS may form a thermal mound.
This thermal mound is a source of black-body photons of emissivity $\qbb$ (dashed four-dotted, brown),
which on
encountering with hot electrons can get Comptonized, whose emissivity is represented by $\qcbb$ (dashed triple-dotted, orange). 
The total emission from all these processes combined, is shown by  solid black line. The flow parameters are same as that in Fig. \ref{fig:4}. 
 We find that  $\qbr$ dominates in the pre-shock region.
As soon as the flow encounters primary shock, emission from 
all the processes increase due to the increase in number density of the system (see, Fig. \ref{fig:4}d). After the primary shock, 
$\qcsy$ dominates over all other emission processes. But very near the surface,
$\qbr$ and $\qcbr$ dominate  because of the rapid increase in $n$. 
Also in this region, there is  reduction in synchrotron  and its Comptonized emission because of the  decrease in $\te$ due to the 
increased cooling. The total luminosity of this system is 
$L=2.759\times 10^{34}$erg s$^{-1}$. 
Contribution of individual  emission processes to the 
total luminosity are: bremsstrahlung=$6.909\times 10^{32}$ erg s$^{-1}$, synchrotron=$5.567 \times 10^{30}$erg
s$^{-1}$, Comptonized synchrotron=$1.589 \times 10^{34}$erg s$^{-1}$, Comptonized bremsstrahlung=$1.100 \times 10^{34}$erg
s$^{-1}$ and Comptonized blackbody=$2.052 \times 10^{30}$erg s$^{-1}$.

In Fig.~\ref{fig:5}b, we represent the spectrum in terms of $L_\nu/(h\nu)$ (in units of s$^{-1}$ keV$^{-1}$) vs $h\nu$ (in keV).
It is apparent from the figure, that bremsstrahlung emission contributes from radio to gamma rays: covering the whole
electromagnetic spectrum, but  it mainly dominates upto near-infrared frequencies. A hump in optical and near UV regime is 
contributed  by synchrotron emission. A power law of spectral index $\alpha=0.365$, covering from UV to X-rays is because of Comptonized synchrotron. A second hump is formed in the gamma ray region which is contributed by Comptonized bremsstrahlung. 
Blackbody emission from the thermal mound at the NS surface is represented by dashed four-dotted brown line and contributes
mainly in the low energy part, while its Comptonized component represented by dashed triple-dotted orange line, contributes
in the UV and soft X-rays.



\subsection{Contribution of different regions of an accretion flow in the emitted spectrum}

\begin{figure}
\centering
\hspace{0.0cm}
\includegraphics[width=9.cm,trim={0.cm 0.7cm 1.0cm 0.5cm},clip]{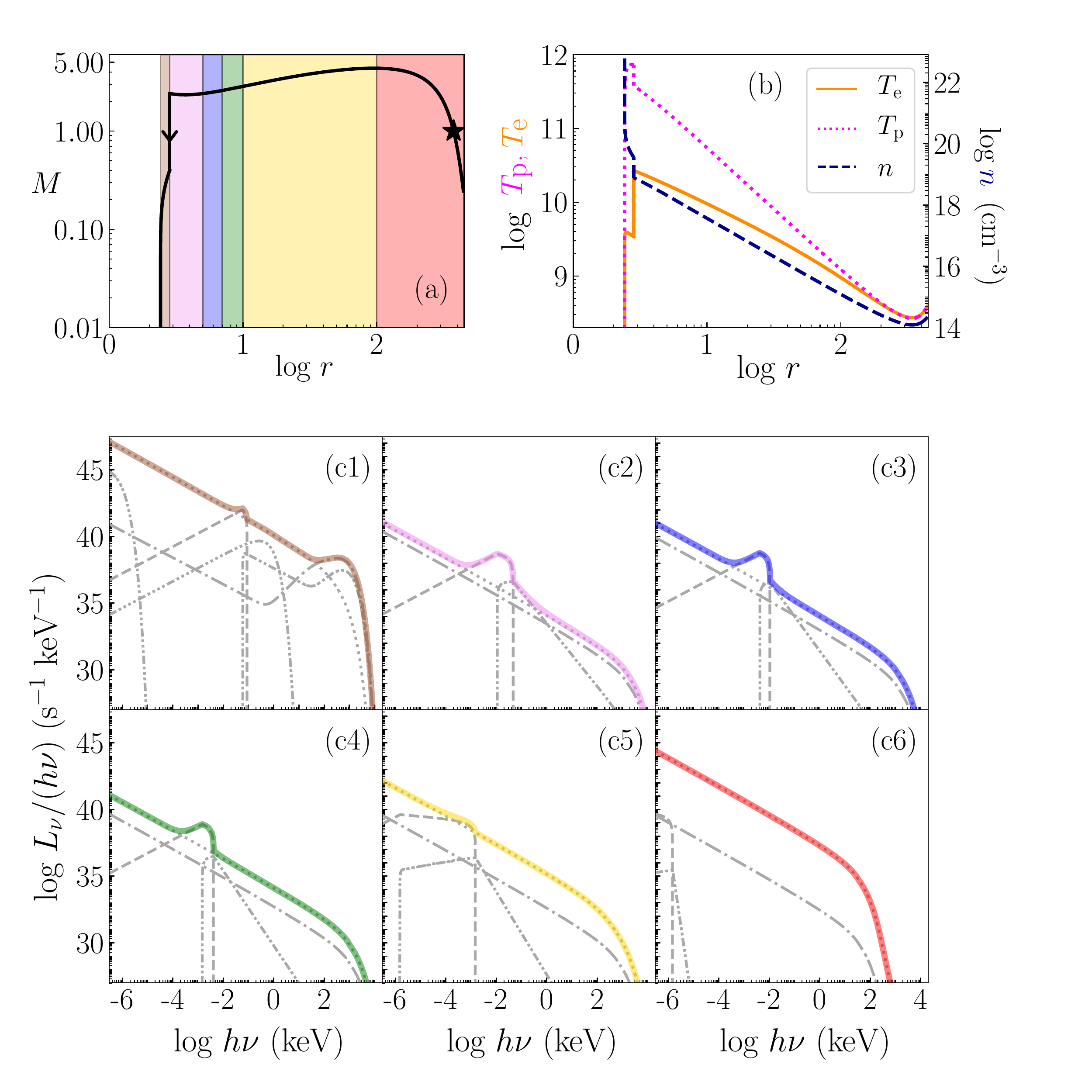}
\vspace{0.0cm}
\caption[] {\small 
(a) Accretion solution (solid, black) is plotted. Different regions have been marked, which are:
$2.418\rg-2.829\rg$ (brown), $2.829\rg-5\rg$ (violet), $5\rg-7\rg$ (blue), 
$7\rg-10\rg$ (green), $10\rg-100\rg$ (yellow) and $100\rg-444.810\rg$ (red). (b) Variation of $\te$ (solid, orange), $\tp$
(dotted, magenta) and $n$ (dashed, dark-blue) 
are plotted. 
Spectral contribution of different regions marked in panel (a) have been correspondingly plotted in solid curves in panels (c1--c6), using the same colour coding. 
In each of these panels, contribution from different emission processes are represented in grey. 
The flow parameters used are, $E = 0.9985$, 
$P = 1.15$s, $\mdot= 10^{15}$g/s and $B_*= 5\times10^9$G.}
\label{fig:6}
\end{figure}

\noindent In Fig.~\ref{fig:6} we examine the contribution of different regions of an NS accretion flow to the observed spectrum.
The flow parameters are $E = 0.9985$, $P = 1.15$s and $\mdot= 10^{15}$g/s, with surface
magnetic field, $B_*= 5\times10^9$G.
In Fig. \ref{fig:6}a, we plot $M$ vs log $r$. The global accretion solution passes through $\rcou=375.720\rg$ (black star), which encounters
a primary shock at $\rsp=2.829\rg$ (downward black arrow) which is of strength $\shokst=M_-/M_+=6.082$ and compression ratio CR$_{\rm ps}=n_+/n_-=4.571$.
Profile of $\te$ (solid, orange) and $\tp$ (dotted, magenta) are plotted in Fig.~\ref{fig:6}b left Y-axis and $n$ (dashed, dark-blue) is plotted in right Y-axis. In Fig. \ref{fig:6}a, we have divided the solution into six different regions (shaded with different colours). The regions are as
follows: $2.418\rg (=r_*)$---$2.829\rg$ (brown), $2.829\rg$---$5\rg$ (violet), $5\rg$---$7\rg$ (blue), $7\rg$---$10\rg$ (green), $10\rg$---$100\rg$
(yellow) and $100\rg$--- $444.810\rg$ (red).  The spectrum from these regions are plotted respectively in Figs.~\ref{fig:6}c1--c6 
following the same colour coding, in solid curves. In each of these panels, spectral contribution from different radiative 
processes are plotted in grey: $\qbr$ (dotted), $\qsyn$ (dashed), $\qcbr$ (dashed single-dotted), $\qcsy$ (dashed double-dotted),
$\qbb$ (dashed four-dotted) and $\qcbb$ (dashed triple-dotted). The total bolometric luminosity of the system is
$L=2.376 \times 10^{35}$ erg s$^{-1}$. 
We have tabulated the contributions of each region in table \ref{table:1}.

The existence of an NS hard surface causes the flow to come to a halt at $r_*$. This, combined with the effect of number density jump at $\rps$,
increases the cooling processes to such an extent, that most of the radiation comes from the post-shock region (Fig.~\ref{fig:6}c1).
At $\rps$, temperatures should generally increase due to compression of matter at the shock front, but the excessive radiative cooling causes $\te$ to drop to a lower value. Unlike electrons, protons cannot radiate efficiently. Thus, $\tp$ follows the general trend as expected in a shock transition and jumps  to a higher value. 
Conforming to the above arguments, we find that the post-shock region 
is very bright, with $99.99985\%$ of the total luminosity ($L_{\rm tot}$) contributed from this region. 
Contribution from Comptonized components decrease and becomes negligible for regions $>10\rg$. Similarly, the hump which is a signature of 
synchrotron emission, decrease and vanish for  $r>100\rg$. This is due to the decrease in $B$ with increase in radius (Eq.~\ref{eq:bp}). Also, synchrotron self-absorption peak frequency ($\nu_{\rm t}$), decrease to lower energies, shifting from EUV  in panel (c1) 
to  optical in panel (c4). In regions  $>10\rg$, the total emission is mainly contributed from bremsstrahlung. It may be noted that, since
 we are adding up large sections of accretion flow in the last two panels c5: $10\rg-100\rg$ and c6: $100\rg-444.810\rg$, the amount of 
emission is higher compared to regions presented in panels c3 and c4.


%

\begin{table}
\caption{{Radiative properties of the regions shaded in Fig. \ref{fig:6}}}
\label{table:1}
\centering
\begin{tabular}{c c c c c}
\hline\hline
Panel No.&Colour &Region (in $\rg$) &  $L$ (erg s$^{-1}$) &$\%$ of $L_{\rm tot}$ \\
\hline
c1 & Brown&2.418 -- 2.829 &  2.376 $\times 10^{35}$ &99.99985
\\
c2 & Violet&2.829 -- 5.0 &  1.158 $\times 10^{28}$ &  4.875$\times 10^{-6}$
\\
c3 & Blue&5.0 -- 7.0 &   5.567 $\times 10^{27}$ &  2.343$\times 10^{-6}$
\\
c4 & Green&7.0 -- 10.0 & 4.796 $\times 10^{27}$ &  2.018$\times 10^{-6}$
\\
c5 & Yellow&10.0 -- 100.0 & 1.586 $\times 10^{28}$ &  6.676$\times 10^{-6}$
\\
c6 & Red&100.0 -- 444.810 & 3.221 $\times 10^{29}$ &  1.355$\times 10^{-4}$ 
\\
\hline
\end{tabular}
\end{table}


\subsection{Shock analysis}
In this section, we examine the properties of primary as well as secondary shocks and their spectral signatures. 
\subsubsection{Properties of primary shock}
\label{sec:prim}
\noindent A primary shock is necessary to slow down the matter such that the surface boundary conditions can be satisfied. This slowing down happens because of dissipative or cooling processes, which in turn is responsible for most of the emission coming from an NS. In the last section, we concluded that majority of the emission comes from the post-shock region.  
\begin{figure}
\centering
\hspace{0.0cm}
\includegraphics[width=8.5cm,trim={1.9cm 5.96cm 4.4cm 6.5cm},clip]{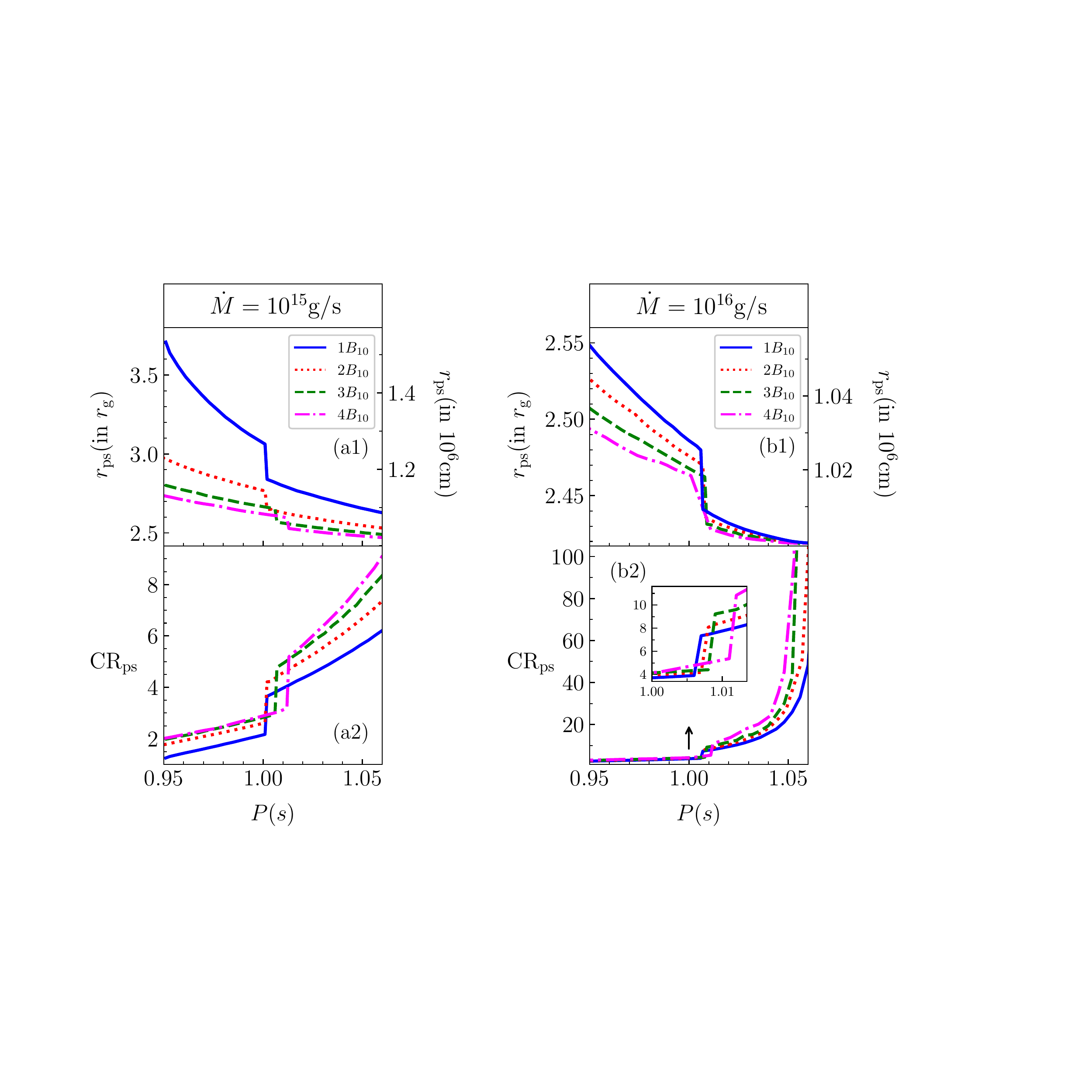}
\vspace{0.0cm}
\caption[] {\small Primary shock location $\rsp$ (a1, b1) and compression ratio CR$_{\rm ps}$ (a2, b2) as  function of  $P$
are plotted for two different $\mdot$: $10^{15}$g/s (a1, a2) and $10^{16}$g/s
(b1, b2). In each panel there are four curves, representing four $B_*$ values, where $B_{10}$ represents magnetic 
field in units of $10^{10}$G.  
$E=0.9984$ is fixed for all cases. }
\label{fig:8}
\end{figure}
We now investigate the variation of primary shock location ($\rsp$) (Figs.~\ref{fig:8}a1, b1) and compression ratio (CR$_{\rm ps}$) (Figs.~\ref{fig:8}a2, b2) with
$P$ as well as $B_*$  of the NS. 
We used four different values of $B_*$: $B_{10}$ (solid, blue), $2B_{10}$ (dotted, red), $3B_{10}$ (dashed, green) and $4B_{10}$ (dashed dotted, magenta), where $B_{10}$ represents magnetic 
field in units of $10^{10}$G. We also use two different values of accretion rate for this purpose, $\mdot=10^{15}$g/s (Figs.~\ref{fig:8}a1, a2) and $\mdot=10^{16}$g/s (Figs.~\ref{fig:8}b1, b2). The specific energy has been fixed to $0.9984$ for all cases. %

Shorter $P$ implies faster spin, so the rotational energy enhances the resistance to the inflowing supersonic
matter. As a result $\rsp$  form at  larger distances. This 
trend is irrespective of the magnitude of $B_*$ and $\mdot$, and is apparent from Figs.~\ref{fig:8}a1, b1. In these plots, the left Y-axis 
represents the shock location in terms of $\rg$ and the right Y-axis represents it in terms of $r_*=10^6$cm, which is the NS radius. At larger distance from the NS surface,
the thermal energy and therefore the pressure is lower. 
As a result, CR$_{\rm ps}$ is lower too (see, Figs.~\ref{fig:8}a2, b2). 
It may be noted that with the increase in $B_*$,  synchrotron and its Comptonization 
increases which in turn reduces the thermal energy. As a result, for a given $P$, increase in $B_*$ decrease $\rps$ and increase
CR$_{\rm ps}$. 
Because of the same reason, $\rps$  is formed nearer to the NS surface for solutions with higher accretion rate  (see Fig.~\ref{fig:8}b1) as compared to the corresponding lower accretion rate solutions  (see Fig.~\ref{fig:8}a1). Since these primary shocks are formed near to the surface, the CR$_{\rm ps}$s have very high values, especially when the rotation period is higher along with 
$B_*$ value. This is seen in Fig.~\ref{fig:8}b2.

In Figs.~\ref{fig:8}a1, b1, a2, b2, as the rotation period is increased, a sudden drop appears near $P\sim 1.0$s. This is due to the change in topology of global  accretion solutions as we change the rotation period. For lower rotation periods (fast spin), matter becomes supersonic on passing through $\rci$, while for higher rotation periods, matter crosses  $\rcou$ and become supersonic. The period $P$ where this transition of accretion flow  occurs from $\rci$ to $\rcou$, $\rsp$ is found to drop to a lower value. The location of this drop depends on  the combination of flow parameters used.

\subsubsection{Properties of secondary shock}
\label{sec:propsec}
\begin{figure}
\centering
\hspace{0.0cm}
\includegraphics[width=8.8cm,trim={.5cm 2.cm 0.cm 7.cm},clip]{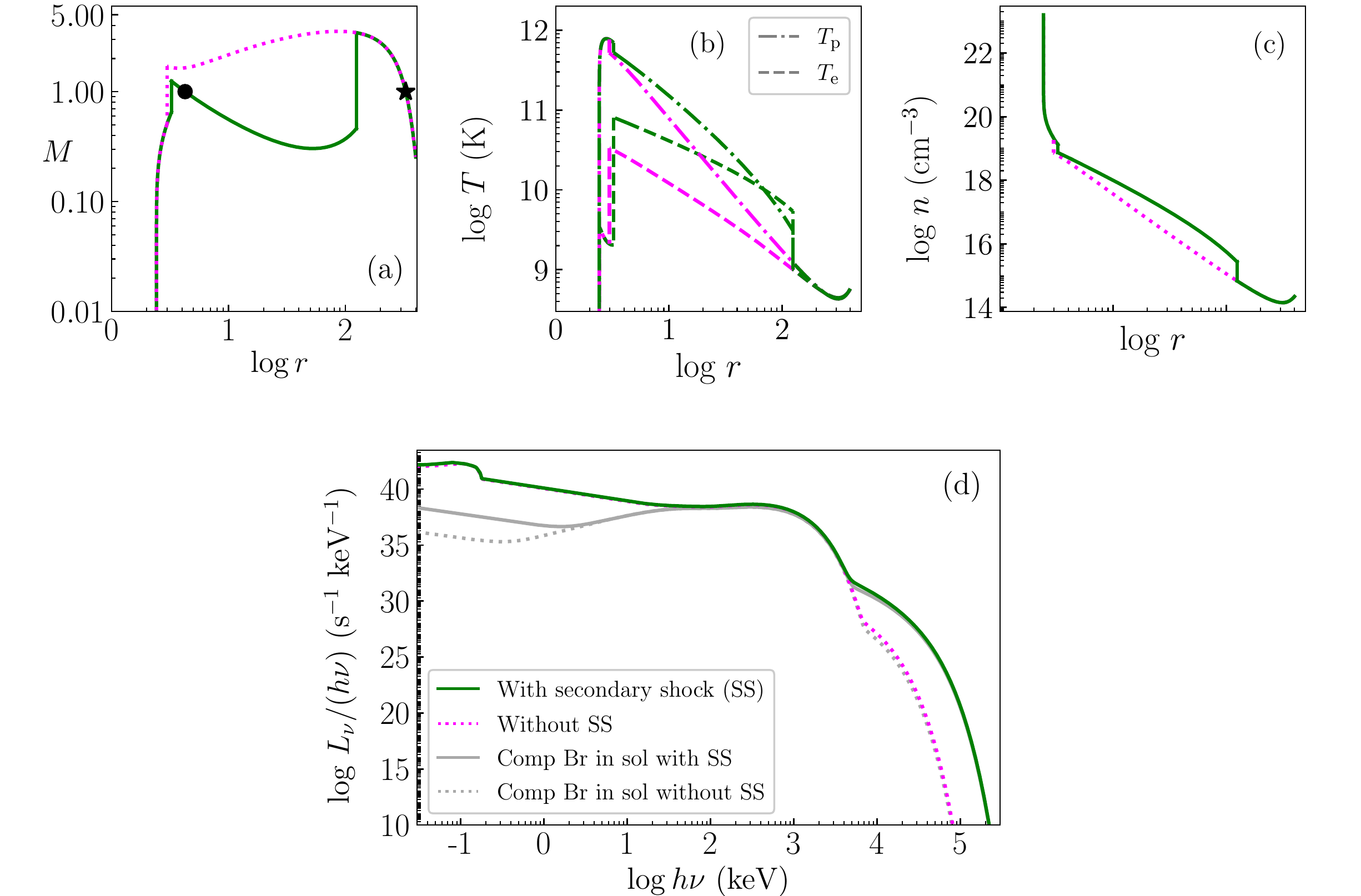}
\vspace{0.0cm}
\caption[] {\small 
Comparison of accretion solutions, with (solid, green) and without secondary shock transition (dotted, magenta) . Plotted in panel (a) $M$ vs $\log r$, where solid black circle and black star 
represents $\rci$ and $\rcou$ respectively; 
(b) $\tp$ (dashed dotted) and $\te$ (dashed) (green is for solution with two shocks, magenta is for solution harbouring only a primary shock); (c) $\log n$; (d) comparison of continuum spectra along with the Comptonized bremsstrahlung components plotted in grey, with the same linestyle as the solutions 
The flow parameters are, $E = 0.9984$, $P = 0.98$s, $\mdot= 10^{15}$g/s and $B_*= 10^{10}$G. 
}
\label{fig:9}
\end{figure}

Now we study the importance of secondary shock in accretion flows around NS and discuss the necessity to obtain a global transonic solution which connects the matter from the accretion disc to NS poles. We present in Fig.~\ref{fig:9}a an accretion solution  along with its other flow variables in Fig.~\ref{fig:9}b ($\te$ and $\tp$) and Fig.~\ref{fig:9}c ($n$). We also plot the spectrum in Fig.~ \ref{fig:9}d. The flow parameters are, $E = 0.9984$, $P = 0.98$s, $\mdot= 10^{15}$g/s and $B_*= 10^{10}$G.
The solution passes through $\rcou$ (marked using a black star) and becomes supersonic, which is shown in Fig. \ref{fig:9}a. Centrifugal and pressure gradient forces oppose this supersonic
matter which drives a shock at $\rss=124.074\rg$. This is termed as a secondary shock (SS). The post-shock subsonic flow then picks up speed
due to the gravity of NS and again becomes supersonic at $\rci$ (black circle). This supersonic flow finally encounters the hard
surface of the NS and as a result settles on it after passing through a primary shock at $\rsp=3.247\rg$. Accretion solution
which passes through these two shock locations: $\rsp$ and $\rss$, is presented using a solid, green curve in Fig. \ref{fig:9}a.
Suppose, we do not check for the Rankine-Hugoniot shock conditions (Eqs.~ \ref{eq:shockcond1}-- \ref{eq:shockcond3}) for regions $r<\rcou$, then the flow will remain supersonic until it reaches near the NS surface, where it will encounter only a primary shock at $\rsp=2.976\rg$. 
This flow is represented using dotted, magenta line. 
It is clear that the two solutions
are quite different. In Fig. \ref{fig:9}b, we compare the corresponding temperature distributions of both the electron
population $\te$ (dashed) and the proton population $\tp$ (dashed dotted) of the accretion solution with two shocks (green) and
with only a single primary shock (magenta). Solution which does not harbour the secondary shock (dotted, magenta) is found to be much colder. Hence, $\rps$ for this solution is formed closer to the NS surface, where the thermal pressure is large. $\te$ at $\rsp$ for both the solutions decreases, while at  $\rss$, it increases.
This is because, in the post-shock flow of $\rss$, the velocity start to increase just after the initial downward jump
at the shock front. This reduces the infall time scales, thereby  prohibiting 
the post-shock flow to loose enough
 energy through cooling. In case of primary shock, the velocity decrease at the shock front similar to $\rss$, but it also steadily decrease afterwards until it reaches the NS surface, where it finally settles down. 
Infall timescales in these cases become larger, allowing matter to radiate more. Therefore after  $\rsp$, $\te$ gradually decreases.
 The radiative processes are not significant in case of proton gas, and thus its temperature increases in the post-shock
region irrespective of whether it is a primary or secondary shock. From the number density distribution in Fig. \ref{fig:9}c,
it is clear that the solution with two shocks (solid, green) is denser compared to the solution with only a
primary shock (dotted, magenta). 
In Fig. \ref{fig:9}d, we compare the continuum spectra of accretion flows with two-shocks (solid, green) and the solution with
only primary shock (dotted, magenta). We plot in grey, the corresponding Comptonized bremsstrahlung components, keeping the same linestyle. 
The solution which harbours secondary shock as argued before, is  hotter and much denser. Apart from that, the secondary shock is also quite strong. Because of all these reasons,  the high energy tail of the spectrum extends beyond $10^5$keV, most of the emission being  contributed from the enhanced Comptonized bremsstrahlung post $\rss$.
This extra high energy component obtained in gamma rays is a signature that a secondary shock exist. 

%
%
%
%

\begin{figure}
\centering
\hspace{0.0cm}
\includegraphics[width=8.5cm,trim={1.9cm 5.96cm 4.4cm 6.5cm},clip]{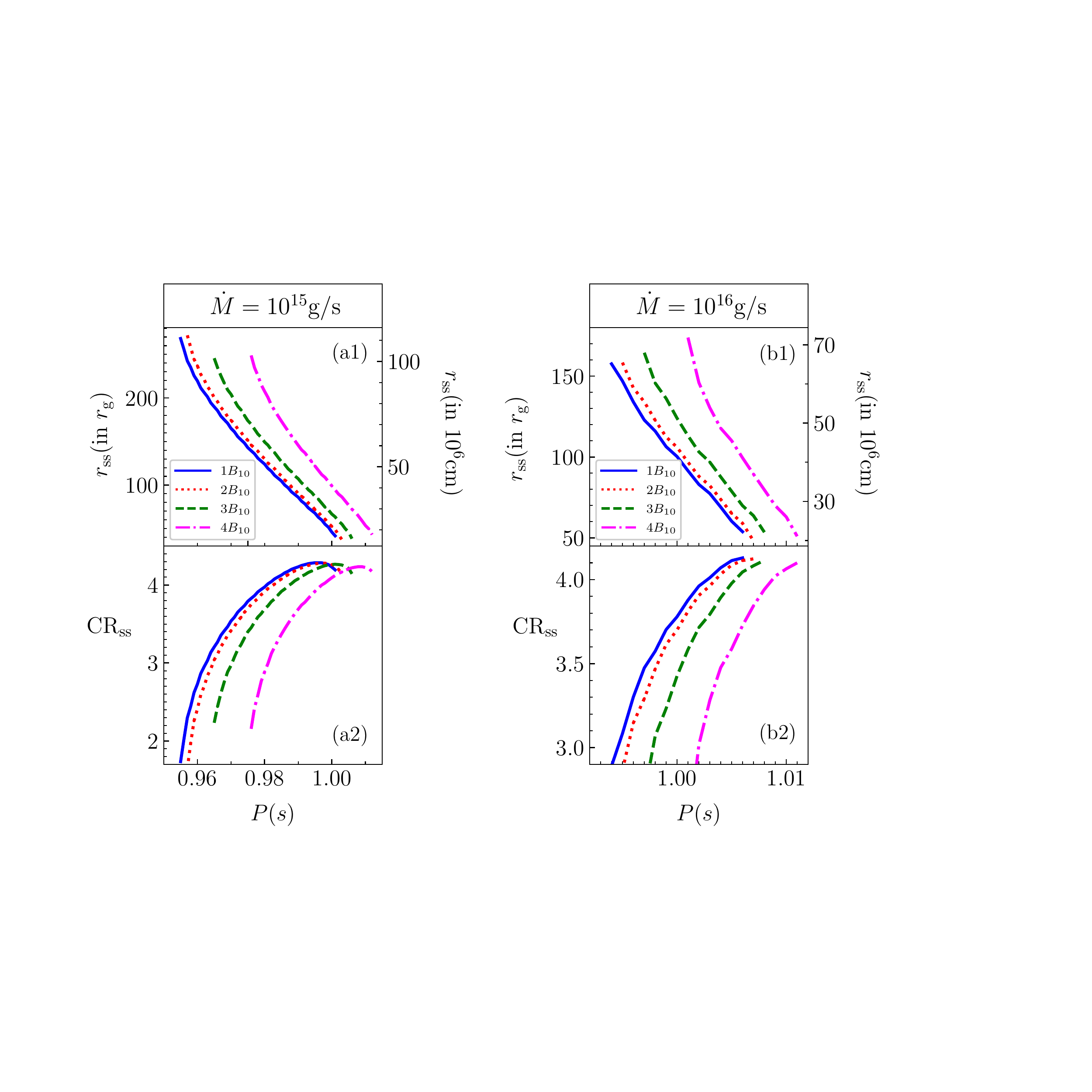}
\vspace{0.0cm}
\caption[] {\small Secondary shock location $\rss$ (a1, b1) and compression ratio CR$_{\rm ss}$ (a2, b2) as  function of period $P$
are plotted for two different accretion rates $10^{15}$g/s (a1, a2) and $10^{16}$g/s
(b1, b2). In each panel there are four curves, representing four $B_*$ values, where $B_{10}$ represents magnetic 
field in units of $10^{10}$G.  
$E=0.9984$ for all cases. }
\label{fig:10}
\end{figure}
In Fig.~\ref{fig:10} we did a similar study as Fig.~\ref{fig:8} but this time it is for secondary shocks. 
Unlike 
primary shocks, secondary shocks are formed for a certain combination of flow parameters and are generally located  far away from the star's 
surface, as is apparent from the values presented in Figs.~\ref{fig:10}a1, b1.
For any given $B_*$, the shock location decreases with increasing $P$ (slow spinning) and therefore the compression ratio CR$_{\rm ss}$ increases. This is similar to what was observed in case of primary shocks (see, Fig.\ref{fig:8}a1, b1). The centrifugal force decreases, which causes the 
shock to move towards the  surface. 
For a given value of $P$, $\rss$ increase with $B_*$, thereby decreasing CR$_{\rm ss}$. Similar effect occurs when the accretion rate is increased. 
It may be noted that even though compression ratio of the secondary shock is less than the primary shock, but the secondary shock by itself is quite strong. 


\subsection{Effect of magnetic field and spin period on the  solution }
\label{subsec:magper}
\begin{figure}
\centering
\hspace{0.0cm}
\includegraphics[width=9.0cm,trim={0.cm 1.5cm 0.6cm 4.5cm},clip]{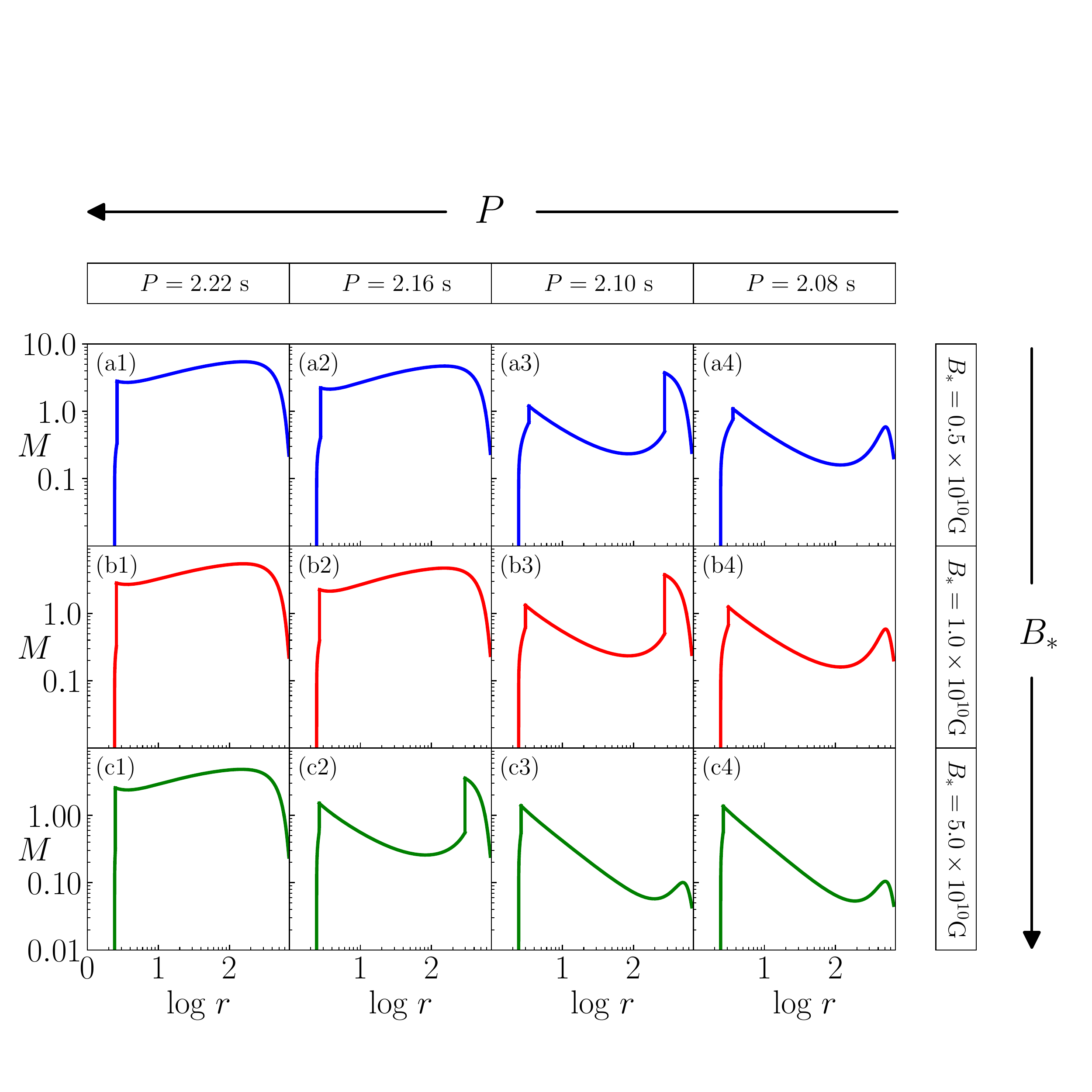} 
\vspace{0.0cm}
\caption[] {\small 
Plot showing the variation of solutions with change in $B_*$ (top to bottom) and  $P$ (left to right) of
the NS. 
Other flow parameters  are $E=0.999$ and $\mdot=10^{15}$g/s.}
\label{fig:11}
\end{figure}
\begin{figure}
\centering
\hspace{0.0cm}
\includegraphics[width=9.0cm,trim={0.cm 1.5cm 0.6cm 4.5cm},clip]{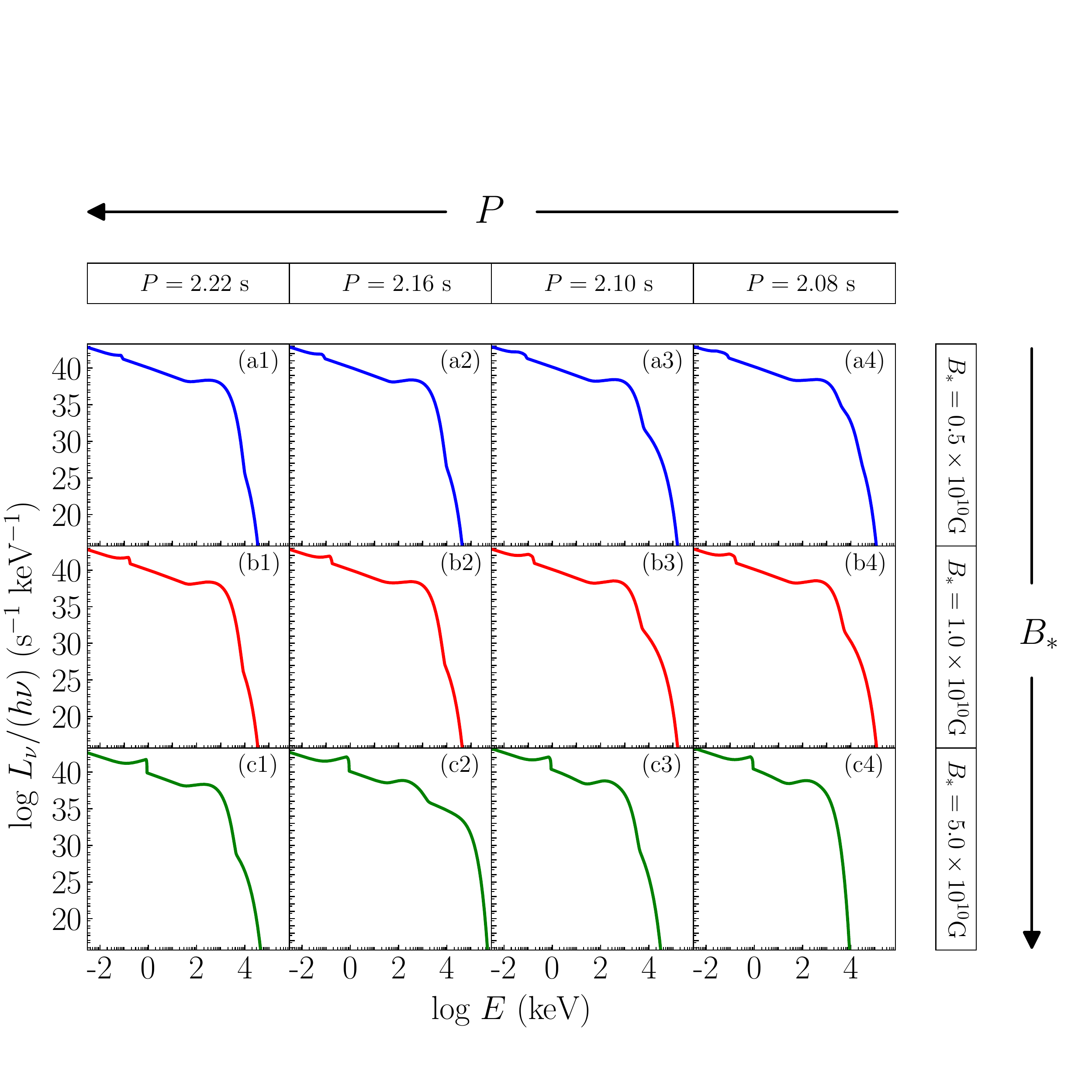} 
\vspace{0.0cm}
\caption[] {\small Plot showing the corresponding change in observable spectra with change in $B_*$ and 
$P$ of the NS. Parameters used are same as in Fig. \ref{fig:11}.}
\label{fig:12}
\end{figure}
In this section, we study the change in solution topology and the observable spectra in Figs.~\ref{fig:11} and \ref{fig:12} respectively for varying 
 surface magnetic field values ($B_*$) and rotation periods ($P$) of the NS system. In both the figures, $P$ 
decrease from left to right with values $2.22$s, $2.16$s, $2.10$s and $2.08$s, labelled as 1--4, while $B_*$  increase on going from top to the bottom, with values: $0.5\times 10^{10}$G, $1.0\times 10^{10}$G and $5.0\times 10^{10}$G and labelled as a--c. Rest of the parameters
are $E=0.999$ and $\mdot=10^{15}$g/s. 
For high spin period or slowly rotating magnetized star, e.g.  $P=2.22$s (Figs.~\ref{fig:11}a1, b1, c1), rotational energy is low, therefore the accretion solution
possess only one outer type sonic point or $\rcou$. However, for fast rotating NS or spin period like $P=2.08$s (Figs.~\ref{fig:11}a4, b4, c4), the rotational
energy is quite high, such that the accretion flow can become supersonic ($M>1$) only very close to the star's surface, i.e., possess
only $\rci$. For intermediate values of $P$, 
 solutions would  harbour secondary shocks (Figs.~\ref{fig:11}a3, b3, c2). 

Fixing $B_*$, if $P$ is varied, we see that for low values of $B_*$ like $0.5\times 10^{10}$G and $1.0\times 10^{10}$G, secondary shock can be formed for fast rotating NS, $P=2.10$s, where 
centrifugal force is large (see Figs. \ref{fig:11}a3, b3). However for higher magnetic fields $B_*=5\times 10^{10}$G, they can be found even for slow rotating 
NS, $P=2.16$s (Fig. \ref{fig:11}c2).  This is because, $B$ determines the flow geometry or in other words, the flux tube thickness. This  affects  $n$, which in turn influence  $v$ and $\te$. Also, the synchrotron cooling and its Comptonization depends on the value of $B$.
Redistribution of the flow variables 
 trigger  shock transition. 

The corresponding continuum spectra for the flows presented in Fig.~\ref{fig:11} are given in Fig. \ref{fig:12}.
Bolometric luminosity ($L$) increases with decrease in $P$. A fast spinning NS would restrict the infalling matter, allowing it to radiate for a longer duration. But with increase in $B_*$ for a given $P$, $L$
decreases. It may be remembered that $>99\%$ of total luminosity comes from the region between $r_*$ and $\rsp$
and for higher $B_*$, post-shock $\te$ is smaller. Therefore, the total luminosity is less for flows with higher $B_*$.
Moreover, flows with higher $B_*$ produce a  prominent synchrotron self-absorption peak, which  shift to higher energies with increase in $B_*$.
Interestingly, the combination of $P$ and $B_*$ which admits secondary shocks (Figs. \ref{fig:11}a3, b3, c2; and \ref{fig:12}a3, b3, c2), exhibits extended high energy tail and high energy cut-offs ($>10^5$keV) similar to Fig.~\ref{fig:9}d. Extensive discussion about this feature has already been done in Section \ref{sec:propsec}.

\begin{figure}
\centering
\hspace{0.8cm}
\includegraphics[width=6.5cm,trim={9.cm 10.cm 6.cm 6.cm},clip]{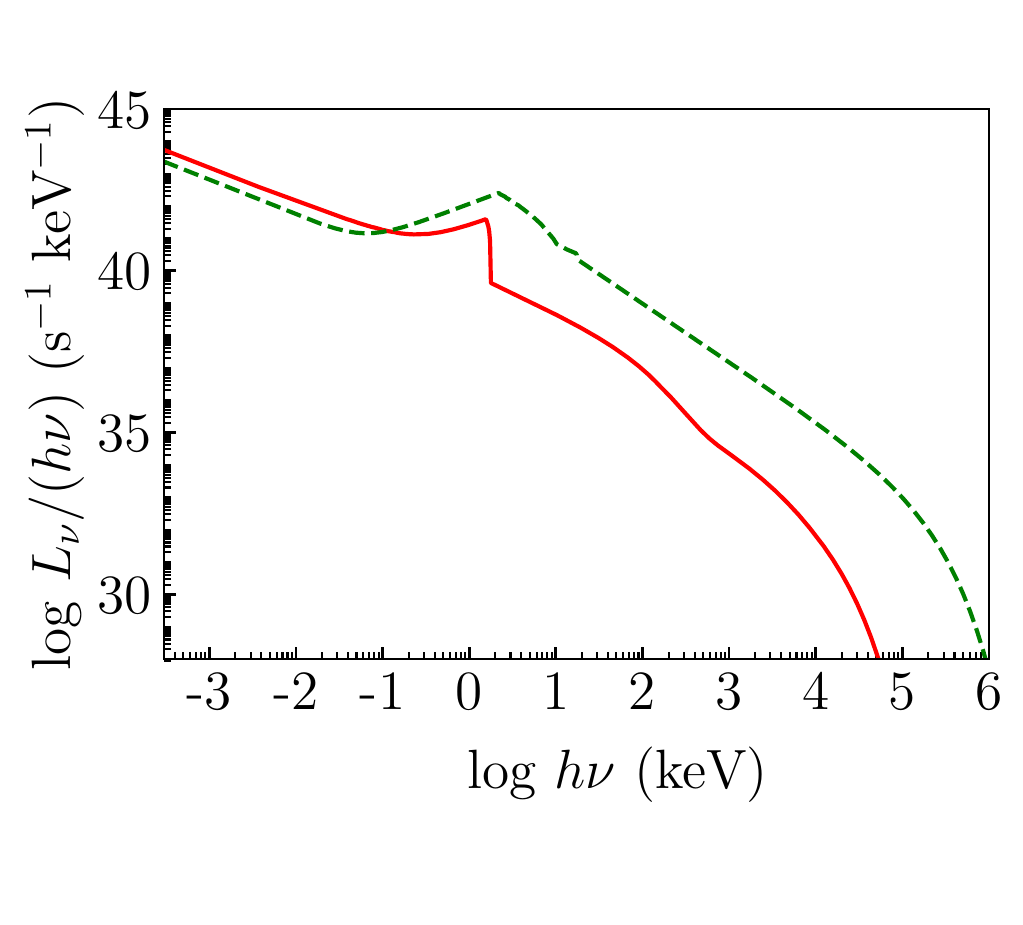} 
\vspace{-0.8cm}
\caption[] {\small Comparing the spectra for two surface magnetic field configuration $B_*= 10^{11}$G (solid, red) and $10^{12}$G (dashed, green). 
Flow parameters are, $P = 5.1$s, $\mdot= 10^{15}$g/s and $E=0.999$.}
\label{fig:13}
\end{figure}

{It may be noted that we did not consider very high magnetic field ($\ge 10^{11}$ G) in Figs. \ref{fig:11} \& \ref{fig:12}. 
Recently \cite{kl15} inferred a dipole magnetic ﬁeld strength of $\sim 10^9$ G for M82 X-2. Also, \cite{king17, kl19,kl20} inferred from their model that pulsating ultra-luminous X-ray sources having NSs as their central objects, have dipole magnetic ﬁeld strengths of the order of $10^9-10^{13}$ G, with majority falling between $10^{10
}$ and $10^{11}$ G \citep{da21}. The magnetic field ranges in these systems are similar to most of the cases that have been discussed in the current work. 
To briefly see the effect of higher magnetic field on accretion flows we consider two NSs, one with surface magnetic field $10^{11}$ G and another with $10^{12}$ G and plot their spectrum in Fig. \ref{fig:13}, keeping all other flow parameters same. As expected, higher surface magnetic field systems creates a more luminous accretion column with a harder spectra in the higher energies.}


\begin{figure}
\centering
\hspace{0.0cm}
\includegraphics[width=8.5cm,trim={0.5cm 2.2cm 11.cm 0.2cm},clip]{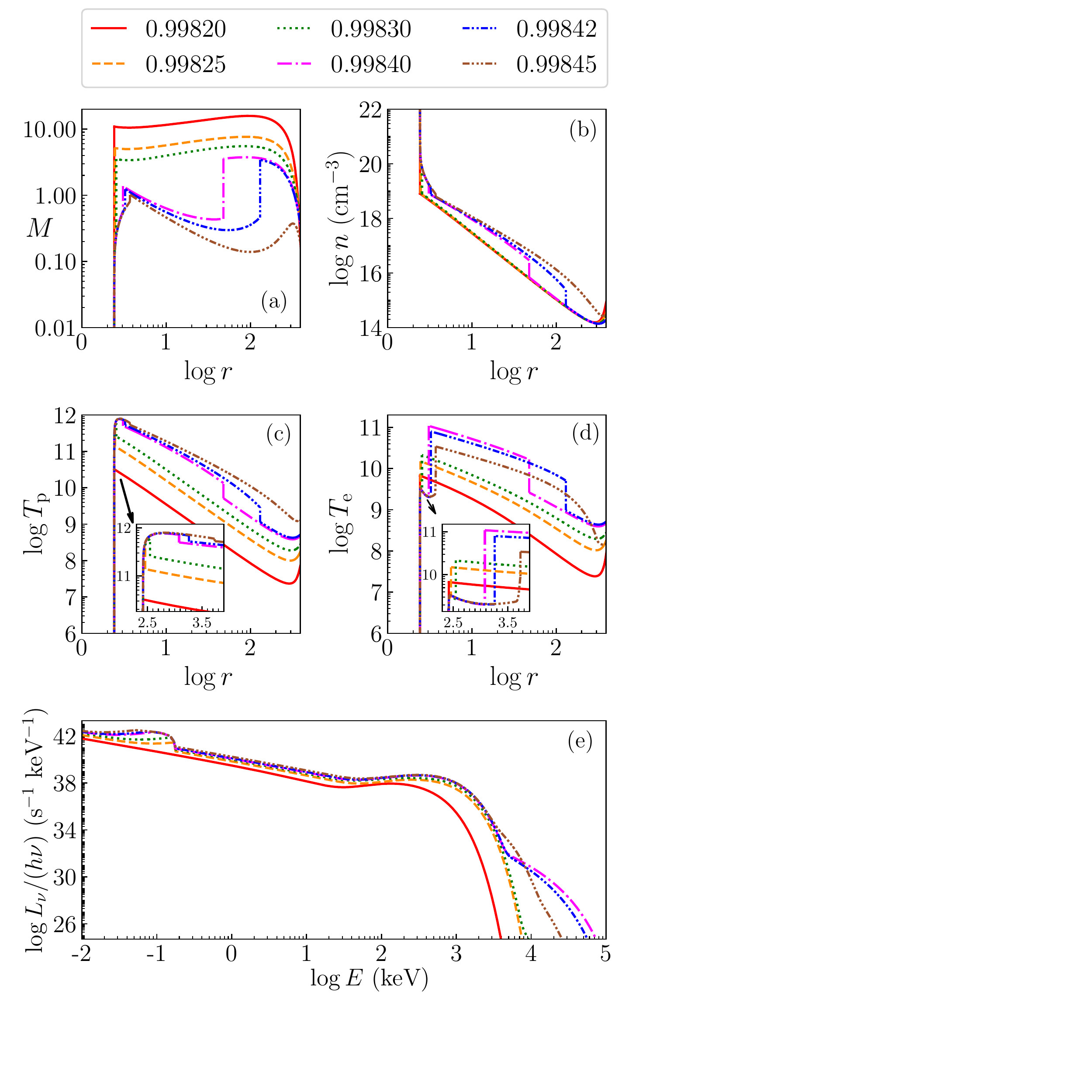} 
\vspace{0.0cm}
\caption[] {\small Effect of variation of $E$ on different  accretion flow variables and  continuum spectra are presented. 
Different values of $E$ are in the legend.
Flow parameters are, $P = 1.0$s, $\mdot= 10^{15}$g/s and $B_*= 10^{10}$G.}
\label{fig:14}
\end{figure}

\subsection{Effect of Bernoulli parameter  ($E$)}

In Fig.~\ref{fig:14}, we analyse the effect of Bernoulli parameter on accretion flows around NS.
We plot in panel (a)  $M$, (b) $\log \tp$, (c) $\log \te$,
(d) $\log n$ and (e) continuum spectra. Each
curve inside these panels, correspond to different values of $E$: $0.99820$ (solid, red), $0.99825$ (dotted, orange), $0.99830$ (dashed, green), $0.9984$ (dashed single-dotted, magenta), $0.99842$
(dashed double-dotted, blue) and $0.99845$ (dashed triple-dotted, brown).
Other flow parameters are, $P = 1.0$s,
$\mdot= 10^{15}$g/s and $B_*= 10^{10}$G. 
Higher $E$ implies higher temperature distribution (i.e., higher $\as$).
Therefore,  matter becomes supersonic ($v>\as$)
after it is accelerated to a much higher infall velocity by gravity, i.e., the sonic point moves inward. This is seen in case of $E=0.99845$ (dashed triple-dotted, brown) where the
accretion flow has single $\rci$.
For low $E$, the accretion flow has one $\rcou$. This is found for  $E=0.99820$ (solid, red), $0.99825$ (dotted, orange) and $0.99830$ (dashed, green). 
For intermediate values of $E$, MCPs may exist, which allows for the formation of secondary shocks ($E=0.9984$, dashed single-dotted, magenta and $E=0.99842$,  dashed double-dotted, blue). $\rss$ will be formed at a large distance for higher $E$, because of higher thermal energy. Generally $\tp$ is higher for higher $E$ (Fig. \ref{fig:14}b), while $\te$ is highest for energies
which harbour secondary shocks (Fig. \ref{fig:14}c).
In Fig.~\ref{fig:14}d, variation in $n$ is presented which shows that it increases with increase in $E$. When an accretion flow harbour secondary shock, there is a distinct density jump seen at the shock front.  
This leads to enhanced cooling, which
is responsible for the appearance of  a prominent
high energy tail (dashed single-dotted, magenta and dashed double-dotted, blue curves in Fig. \ref{fig:14}e). But this does not increase the bolometric luminosity substantially since $\rss \gg\rps$ and it is already known that most of the luminosity is contributed from regions $<\rps$. However, the $L$ 
increase with increase in $E$.
In table \ref{table:2}, we summarize the effect of variation of $E$ on accretion flows around NSs. We list the values of $\rps$ and $\rss$ (if any), their CRs and the corresponding $L$.

\begin{table}
\caption{{The effect of variation of $E$ on solutions plotted in Fig. \ref{fig:14}}}
\label{table:2}
\centering
\begin{tabular}{c c c c c c}
\hline\hline
$E$ &\multicolumn{2}{c}{Primary shock} & \multicolumn{2}{c}{Secondary shock} & $L$ (erg s$^{-1}$)  \\
 &$\rsp$ & CR$_{\rm ps}$ & $\rss$ & CR$_{\rm ss}$ & $\times  10^{35}$ \\
\hline

0.99820& 2.419 & 229.896& --&--& 0.068
\\
0.99825&2.461 & 16.628 & --&--& 0.922
\\
0.99830& 2.548 &  8.736&  --&--& 1.595
\\
0.99840& 3.079 & 2.128& 47.888& 4.244& 2.769
\\
0.99842& 3.260 & 1.748&129.522 &3.950& 2.915 
\\
0.99845& 3.732 & 1.207&  --&--&3.114
\\
\hline
\end{tabular}
\end{table}

\subsection{Effect of variation of accretion rate}

\begin{figure*}
\centering
\hspace{0.0cm}
\includegraphics[width=16.7cm,trim={0.cm 7.2cm 0.cm 2.5cm},clip]{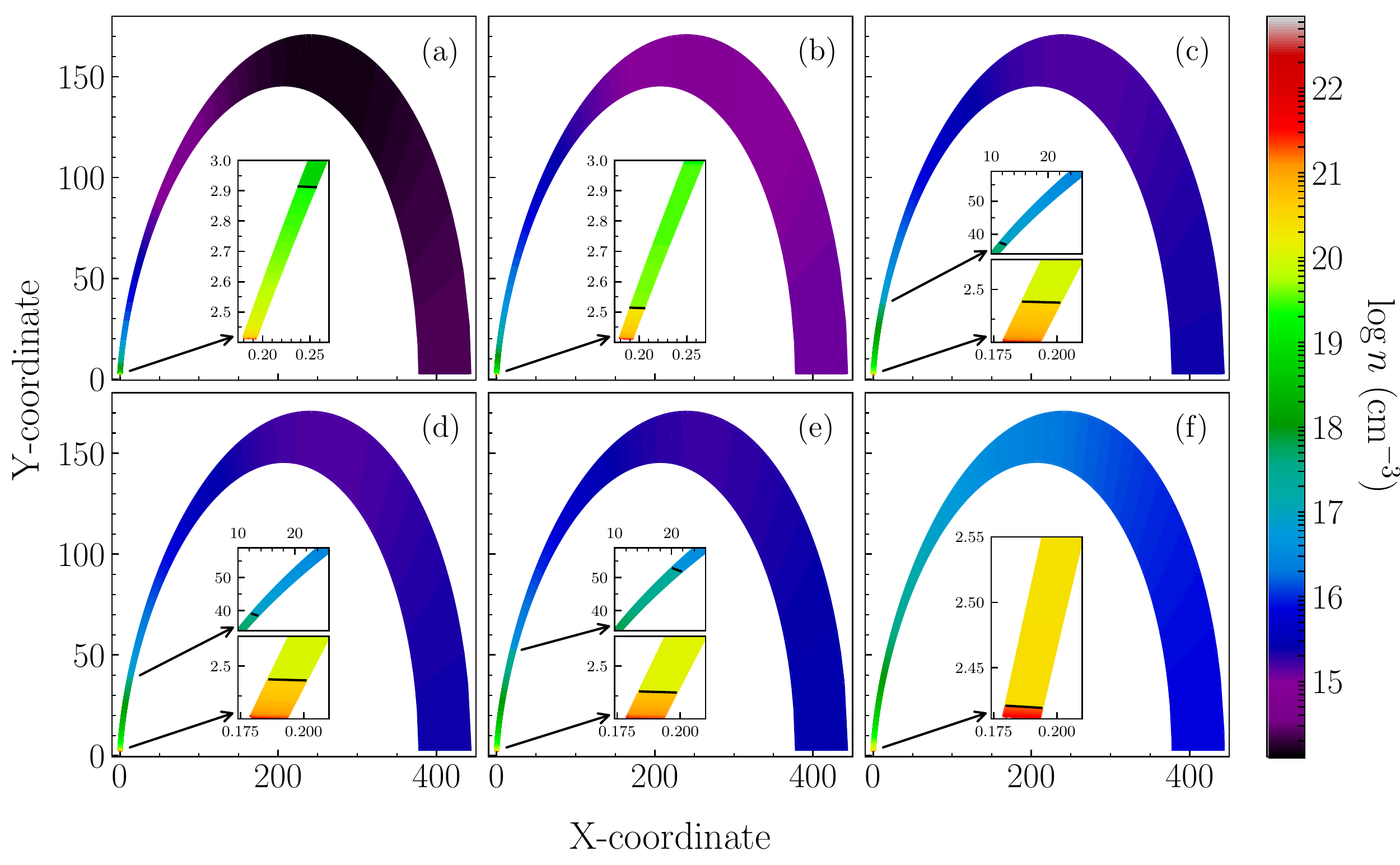} 
\vspace{0.0cm}
\caption[] {\small The variation of
number density ($n$) inside a flux tube represented in two spatial dimension. 
The different panels correspond to systems with different $\mdot$: (a) $1.0\times 10^{15}$g/s, (b) $5\times 10^{15}$g/s, (c) $9.8\times 10^{15}$g/s,
(d) $1.0\times 10^{16}$g/s, (e) $1.2\times 10^{16}$g/s and (f) $2.5\times 10^{16}$g/s. Other parameters are $E=0.9983$, $P=0.91$s
and $B_*=8\times 10^9$G. }
\label{fig:15}
\end{figure*}
In Fig. \ref{fig:15} each panel, we plot the variation in number density ($n$)  along a flux tube represented in two dimensional X-Y plane. The number density values, in units of  cm$^{-3}$, are presented using a colour bar.
The different panels show system with different $\mdot$: (a)
$1.0\times 10^{15}$g/s, (b) $5\times 10^{15}$g/s,
(c) $9.8\times 10^{15}$g/s, (d) $1.0\times 10^{16}$g/s, (e)
$1.2\times 10^{16}$g/s and (f) $2.5\times 10^{16}$g/s.
The other flow parameters are, $E=0.9983$, $P=0.91$s and $B_*=8\times 10^9$G.
$n$ increases with increase in $\mdot$. However, due to strong magnetic field assumption,
the flux tube width is independent of the $\mdot$ value, which is apparent from the figure. 
In Figs. \ref{fig:15}a, b,
the flow passes through one $\rcou$. 
With  increase in $\mdot$, accreting matter becomes hotter, increasing thermal pressure and as a result secondary shock
appear, which is seen in Figs. \ref{fig:15}c--e. In the inset of each panel, we zoomed the region around
primary and secondary shocks. 
The location of secondary shock ($\rss$) for panels c--e are, $38.131\rg,~40.848\rg$ and $56.426\rg$, respectively,
while $\rsp$ values for panels a--f are, $2.928\rg,~2.526\rg,~2.493\rg,~2.491\rg,~2,470\rg$ and $2.428\rg$,
respectively.
If we further increase the accretion rate of the system, then the flow passes through a single inner
sonic point ($\rci$), which is formed very close to the NS surface. This is seen for accretion rate $2.5\times 10^{16}$g/s (Fig. \ref{fig:15}f). All these solutions harbour primary shock irrespective of the presence of any secondary shock. 
%

\begin{figure}
\centering
\hspace{0.0cm}
\includegraphics[width=8.8cm,trim={0.cm 7.2cm 0.cm 2.5cm},clip]{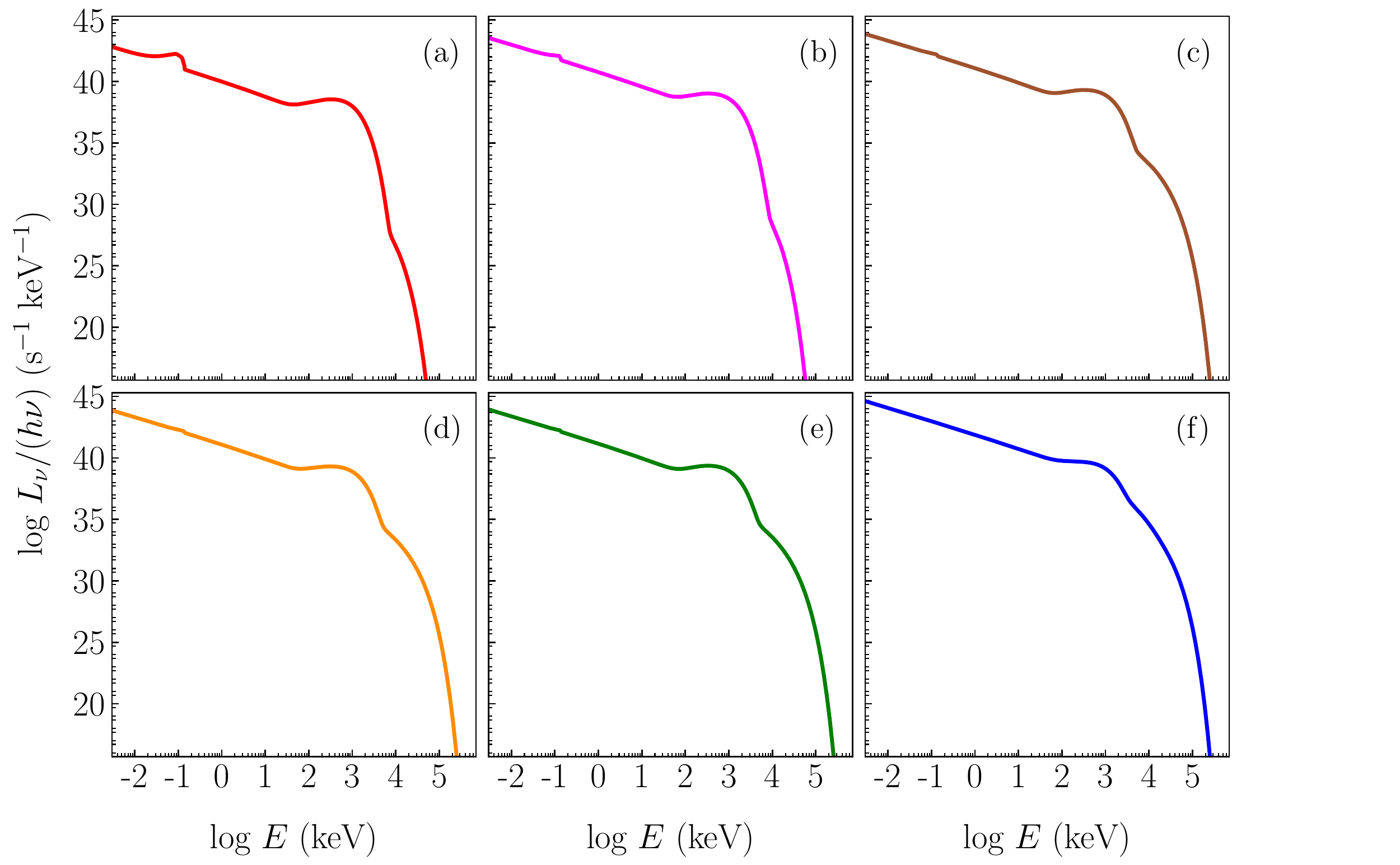} 
\vspace{0.0cm}
\caption[] {\small Spectra corresponding to the flows presented in Fig. \ref{fig:15}.}
\label{fig:16}
\end{figure}

Spectra corresponding to Fig. \ref{fig:15} are plotted in corresponding panels of Fig. \ref{fig:16}.
The luminosity increases with increase in accretion rate of the system.
A prominent synchrotron turnover frequency is observed around $0.1$keV for $\mdot=1.0\times 10^{15}$g/s (see, Fig. \ref{fig:16}a).
Its magnitude and location remains almost the same irrespective of $\mdot$, because, $\nu_{\rm t}$ depends
strongly on the value of $B$. But this signature subsides with increasing $\mdot$  of the system. It is mildly
visible for $5\times 10^{15}$g/s (see, Fig. \ref{fig:16}b). 
For a given magnetic field structure, increasing $\mdot$, increases $n$ and thereby bremsstrahlung
emission and its Comptonization. Therefore, for higher $\mdot$ systems, the synchrotron turn over frequency is masked by the dominance of bremsstrahlung emission.
For solutions harbouring secondary shock, there is an
extended power law, with the cut-off going to higher energies similar to the spectra discussed in sections before.

\section{Discussions and Conclusions}
\label{sec:con}
In this paper we investigated two-temperature accretion flows 
around strongly magnetized stars specifically NSs. 

The major impediment for obtaining self-consistent two-temperature solution is that, the transonic solution is not unique.
The total number of governing equations are less than the total number of flow variables to be computed. This problem
is same as that observed for two-temperature flows around BHs. To conclude, the degeneracy in two-temperature regime is generic in nature and is irrespective of the type of central object. For a given set of constants of motion, infinite transonic solutions are admissible.  
The problem of degeneracy around BHs were solved  by computing the entropy of the flow very close to the 
event horizon and then choosing the solution with the maximum entropy. It may be noted that, for two-temperature flows there is no analytical expression for entropy. 
But an entropy measure form can be derived only very close to  a point where the accreting matter approach free-fall velocity. However, such a  situation is not possible on or outside the surface of a  magnetized star. Therefore, in this paper we proposed a novel method to obtain a unique transonic two-temperature accretion solution around a strongly magnetized star. {By strong field, we only meant that we chose a surface magnetic field of the star, such that the magnetic energy density is much stronger than the gas energy density, so the magnetic fields are not deformed due to the gas motion and the flow remains sub-Alfv\'enic.}
We traced the projected transonic solution, assuming the star to be more compact, to a point $\rin\sim \rg$ such that the infall velocities at this radius is found to approach free-fall
values. Obtaining all possible solutions for a given set of CoM or constants of motion, we  chose the highest
entropy solution following the second law of thermodynamics. This is the  unique solution.
{It may further be noted that the highest entropy solution at $\rin$ will remain the highest entropy solution at $\rd$. Even if
the flow at $\rd$ starts with arbitrary $\te$ and $\tp$ the solution will be time-dependent until it achieve  values which will correspond to those having highest entropy. Somewhat similar argument can be extended for Bondi flows \citep{b52}, where, in presence of all non-transonic solutions as well as a transonic solution, the flow would choose the highest entropy transonic global solution only. }

After proposing a general methodology to constrain degeneracy, we investigated two-temperature accretion solutions around NS for a wide range of parameter space. 
There are accretion solutions which become
supersonic after passing through an outer sonic point, while others become supersonic after passing through the inner sonic point. 
There are even solutions which passes through outer sonic point and suffers a secondary shock, after which it 
becomes
subsonic. Thereafter, this solution becomes supersonic after passing through the inner sonic point. All these solutions have one thing in common i. e.,
all theses solutions ends up with a terminating shock at the NS surface. This shock ensures that the surface boundary conditions are satisfied.
We found that almost the entire radiation is emitted from this post-shock flow. The secondary shock, thus, do not significantly influence the total luminosity, however it is responsible for an additional high energy tail or an extended high energy cut-off. So, there is a need to study the accretions solutions connecting the accretion disc to the NS poles, and not just investigate the radiative property of a freely falling accretion column onto an
NS.

The compression ratio across the primary shocks are  very high and depend on the combination of flow parameters used. 
Secondary shocks, on the other hand, have compression ratios less than that of primary shocks, but strength of these
shocks are not very low.
Apart from adding high energy tail via inverse-Comptonization
of soft photons, such shocks can also aid in particle acceleration. Presence of high density and very low speed, makes the electron temperature distribution decrease at the primary shock location, due to enhanced cooling, contrary to what is expected at a shock front. 

The combination of flow parameters present in an NS is important in determining the flow topology and the observable spectrum. Apart from the thermal Comptonization mechanisms considered in this paper, there could be bulk motion Comptonization present in the system as well \citep{bw05b,bw07}. An order of magnitude estimate of this emission process has been conducted.
And we arrived at the conclusion that the accretion rates we have considered, radiation from the mound would not exert any significant resistance and
would not change the qualitative nature of the solutions presented. A detailed discussion on this process is beyond the scope of this paper and will be dealt with properly in our future works. 


\section*{Data Availability}
The data underlying this article will be shared on reasonable request to the corresponding author.






\bibliographystyle{mnras}
\bibliography{ms_rev2} 




\appendix


\section{Regenerating one-temperature accretion solution around a strongly magnetized star using the new methodology proposed in this paper}
\label{app:1}

\begin{figure}
\centering
\hspace{0.0cm}
\includegraphics[width=8.cm,trim={0.cm 8.5cm 5.cm 3.3cm},clip]{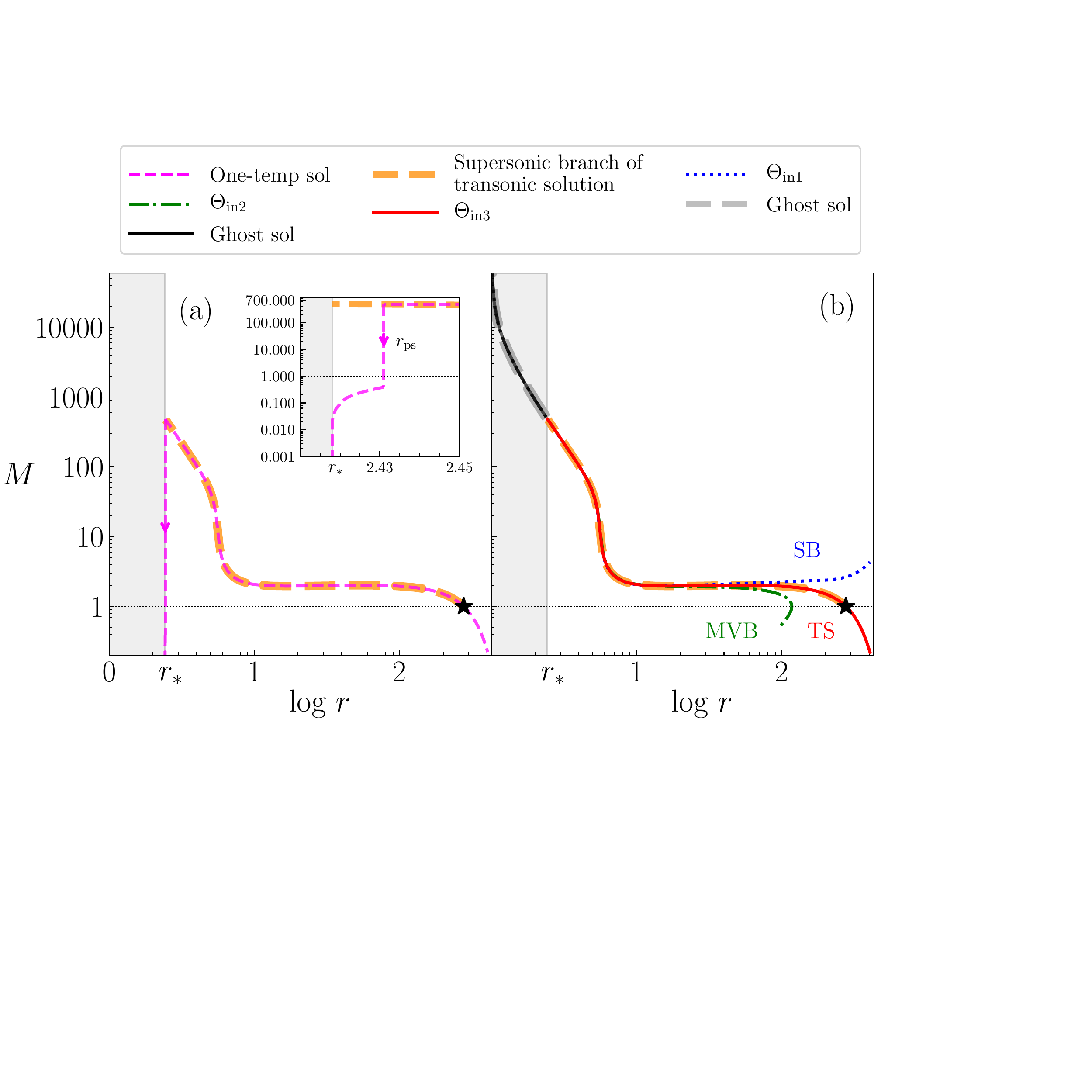}
\vspace{0.0cm}
\caption[] {\small(a) One-temperature solution (dashed, magenta) obtained by sonic point analysis of accretion flows around an NS
\citepalias{sc18b}. The primary shock ($\rsp$) is shown using downward arrow (see zoomed inset). 
Thick dashed orange curve marks just the supersonic branch of this solution. 
(b) Obtaining solutions with $\rin$ as the inner boundary. 
TS (solid, red) obtained completely coincides with dashed magenta/thick dashed orange curve.
The solid black and thick dashed grey lines are part of the TS below $r_*$ and are named as the ghost solutions. The location of primary
shock remains the same. The input parameters used are, $E = 0.9977$,
$P=1.0$s, $\mdot=2.957\times 10^{15}$g/s, $B_* = 10^{10}$G , $M_*=1.4\msol$ and $r_*=10^6$cm$=2.418\rg$.}
\label{fig:a1}
\end{figure}

In the one-temperature regime, for a given set of CoM $E$, $P$
and $\mdot$, a transonic solution 
is unique. 
Therefore, one can solve it using different methodologies but every method will admit a unique solution. 
In this section, we show that our proposed `ghost solution' method 
indeed regenerates the same accretion solution as was obtained by \citetalias{sc18b}, who started the integration from the critical point. 

In Fig. \ref{fig:a1}a, a typical one-temperature transonic accretion solution onto an NS is presented, which has been obtained following the methodology of \citetalias{sc18b}. Here, $M$
is plotted against radial distance ($r$) from the center of the NS. The parameters used are,
$E = 0.9977$, $P=1.0$s, $\mdot=2.957\times 10^{15}$g/s, surface magnetic field $B_* = 10^{10}$G and $M_*=1.4\msol$. Radius of the NS, $r_*=10^6$cm$=2.418\rg$, is marked in the figure and region below it is shaded in grey.
 The global transonic solution (hereafter, abbreviated as TS) (satisfying NS boundary condition) is represented by  dashed magenta curve with the sonic point $\rc$, marked using a black star. The star surface drives a primary shock (downward magenta arrow) at $\rps$,
located just near the surface (see, zoomed inset plot),
after which matter becomes subsonic and then slowly settles down onto the star
($v \rightarrow 0$ at $r \rightarrow r_*$). If the matter would have directly hit the surface of the star without undergoing the primary shock transition, the supersonic branch in such a case
would be the one which is marked using a thick dashed orange line.
The cooling processes considered are same as those present in \citetalias{sc18b}.
The methodology adopted by these authors to obtain an accretion solution is the general `sonic point analysis' method.  
In this method,  given a set of constants of motion, the location of $\rc$ is found first. Then from $\rc$,  the equations of motion are integrated inwards (till the star's surface: to obtain the supersonic branch; thick dashed orange curve) and outwards (till $\rco$: to obtain the subsonic branch). Now, at every radius of the supersonic branch (especially in the region near  the star's surface), they check for the allowed shock transitions until the post-shock branch satisfies the surface boundary condition \citep{dhang21}. In this way, the location of $\rps$ is found and a global TS including surface shock is obtained. 

In Fig. \ref{fig:a1}b, we compare this TS (dashed, magenta) with the `ghost solution method' proposed in this paper. 
We consider a point at $\rin \sim \rg$ where the gravity is very strong such that any other interaction or processes can be considered negligible. Then, for the same set of CoM as before (Fig. \ref{fig:a1}a), we supply a guess value of $\Theta=\Theta_{{\rm in}1}$ at $\rin$. We estimate
$v_{{\rm in}1}~[\equiv \vin(E,\mdot,\Theta_{{\rm in}1})$, equation obtained from the canonical form of Bernoulli parameter] and integrate out from $\rin$, using the EoM. 
Suppose for $\Theta_{{\rm in}1}$ we obtain a completely
supersonic branch (SB, dotted blue) of solution. This branch of solution is not physical. Hence, in the next iteration we change the guess value of $\Theta_{\rm in}$ to $\Theta_{{\rm in}2}$ and correspondingly estimating the value of $v_{\rm in2}$, we  obtain another solution. We might obtain a multi-valued
branch (MVB, dashed dotted green) of solution which too is unphysical. Hence, we iterate in between these two $\thetain$ values until for say $\Theta_{{\rm in}3}$ we get a TS
(solid, red). We can see from Fig. \ref{fig:a1}b, that this solution completely masks the underlying dashed magenta curve, but the thick dashed orange supersonic branch bears
the witness that the two transonic curves have coincided with each other. Portion of the
TS (solid, red) inside the grey shaded region, represented using black curve, is the `ghost solution' or the `projected transonic solution'. This is actually a continuation of the TS obtained by \citetalias{sc18b}, i.e. when the supersonic branch obtained by sonic point analysis method (thick dashed orange) is integrated further inwards till $\rin$ is reached and is not terminated at the NS surface ($r_*$), then the solution obtained below $r_*$ (thick dashed grey)  overlaps with the black curve. The location of primary
shock remains the same for both these methods. Thus, using the `ghost solution' method, we have regenerated the full transonic solution as was obtained by \citetalias{sc18b}.

Hence, we conclude that the new proposed method just uses the property of gravity,
that is, it behaves as if the mass is concentrated at the centre and therefore the `ghost solution' is actually a part of the
full solution, only to be chopped off by the boundary condition.
This method in brief, directs us  to obtain the projected transonic accretion solution, by starting the integration from a region smaller than the actual star surface. 

\bsp	
\label{lastpage}
\end{document}